\documentclass[aps,pre,preprint,onecolumn,citeautoscript,superscriptaddress,floatfix]{revtex4-1}  
\usepackage[utf8]{inputenc}
\usepackage{amssymb,bm,xspace,soul} 
\usepackage{color} 
\usepackage[papersize={8.5in,11in}]{geometry}
\usepackage{natbib}
\usepackage{graphicx}
\usepackage{amsmath}
\usepackage[table]{xcolor}
\usepackage{verbatim}
\usepackage{braket}
\usepackage{enumitem}
\usepackage{feynmp}
\usepackage{math tools} %\Aboxed{ } draws boxes around '&' in align
\usepackage{dsfont} %gives the ``double stroke 1" I like to use for the identity \mathds{}
\usepackage{mathrsfs} %give very curly math script. I use it for \hilb=\mathscr{H}, Hilbert space H 
\usepackage{multirow}
\usepackage{slashed}%gives Dirac slash -- \slashed(\partial) -- I make a shortcut below
\usepackage[colorlinks=true]{hyperref} 
\hypersetup{
    bookmarks=true,         % show bookmarks bar?
    unicode=false,          % non-Latin characters 
    pdftoolbar=true,        % show Acrobat
    pdfmenubar=true,        % show Acrobat 
    pdffitwindow=false,     % window fit to page when opened
    pdfstartview={FitH},    % fits the width of the page to the window
    pdftitle={My title},    % title
    pdfauthor={Author},     % author
    pdfsubject={Subject},   % subject of the document
    pdfcreator={Creator},   % creator of the document
    pdfproducer={Producer}, % producer of the document
    pdfkeywords={keyword1} {key2} {key3}, % list of keywords
    pdfnewwindow=true,      % links in new window
    colorlinks=true,       % false: boxed links; true: colored links
    linkcolor=magenta, %red,          % color of internal links (change box color with linkbordercolor)
    citecolor=blue,        % color of links to bibliography
    filecolor=magenta,      % color of file links
    urlcolor=cyan           % color of external links
} 

\makeatletter
    \def\CT@@do@color{%
      \global\let\CT@do@color\relax
            \@tempdima\wd\z@
            \advance\@tempdima\@tempdimb
            \advance\@tempdima\@tempdimc
    \advance\@tempdimb\tabcolsep
    \advance\@tempdimc\tabcolsep
    \advance\@tempdima2\tabcolsep
            \kern-\@tempdimb
            \leaders\vrule
                    \hskip\@tempdima\@plus  1fill
            \kern-\@tempdimc
            \hskip-\wd\z@ \@plus -1fill }
    \makeatother
\newcommand{\CS}{\mathrm{CS}}
\newcommand{\Dir}{\mathrm{D}}
\newcommand{\Max}{\mathrm{M}}
\newcommand{\FP}{\mathrm{FP}}
\newcommand{\Gg}{\mathrm{G}}

\renewcommand{\O}{\mathcal{O}}

\newcommand{\D}{\mathcal{D}}

\newcommand{\E}{\mathcal{E}}

\newcommand{\vn}{{\mathbf{n}}}

\newcommand{\vq}{{\mathbf{q}}}
\newcommand{\vp}{{\mathbf{p}}}
\newcommand{\vx}{{\mathbf{x}}}

\newcommand{\bpsi}{\bar{\psi}}

\newcommand{\tr}{\text{tr}}
\newcommand{\sd}[1]{\slashed{#1}}

\renewcommand{\o}{\over}
\newcommand{\eq}[1]{\begin{align}#1\end{align}}

\renewcommand{\(}{\left(}
\renewcommand{\)}{\right)}
\renewcommand{\[}{\left[}
\renewcommand{\]}{\right]}
\newcommand{\abs}[1]{\left| #1 \right|}

\newcommand{\nt}{\notag\\}

\let\v\mathbf

\newcommand{\bb}{\boldsymbol}

\newcommand{\ep}{\epsilon}

\renewcommand{\a}{\alpha}
\renewcommand{\b}{\beta}
\renewcommand{\d}{\delta}
\newcommand{\g}{\gamma}
\newcommand{\n}{\nu}
\newcommand{\m}{\mu}
\renewcommand{\t}{\tau}
\newcommand{\s}{\sigma}
\newcommand{\w}{\omega}

\newcommand{\lam}{\lambda}
\renewcommand{\r}{\rho}

\let\ptl\partial
\renewcommand{\dag}{\dagger}

\newcommand{\Zt}{\mathds{Z}_2}
\newcommand{\Z}{\mathds{Z}}
\renewcommand\Re{\operatorname{\mathfrak{Re}}}

\newcommand{\jthetac}[2]{\vartheta\begin{bmatrix}#1\\#2\end{bmatrix}}

\usepackage{tocloft} 
\addtocontents{toc}{\cftpagenumbersoff{chapter}}
\addtocontents{toc}{\cftpagenumbersoff{section}}

\setlist[enumerate]{leftmargin=*}

\begin{document}

\preprint{arXiv:1607.05279}
\title{Spectrum of conformal gauge theories on a torus}
 \author{Alex Thomson}
 \affiliation{Department of Physics, Harvard University, Cambridge, Massachusetts, 02138, USA}
 \author{Subir Sachdev}
 \affiliation{Department of Physics, Harvard University, Cambridge, Massachusetts, 02138, USA}
 \affiliation{Perimeter Institute for Theoretical Physics, Waterloo, Ontario N2L 2Y5, Canada}
 \date{\today}
 \vspace{0.6in}
\begin{abstract} 
Many model quantum spin systems have been proposed to realize critical points or phases
described by 2+1 dimensional conformal gauge theories. On a torus of size $L$ and modular parameter $\tau$, 
the energy levels of such gauge theories equal $(1/L)$ times universal functions of $\tau$. We compute the universal
spectrum of QED$_3$, a U(1) gauge theory with $N_f$ two-component massless Dirac fermions, in the large $N_f$ limit.
We also allow for a Chern-Simons term at level $k$, 
and show how the topological $k$-fold ground state degeneracy in the absence of fermions
transforms into the universal spectrum in the presence of fermions; these computations are performed at fixed $N_f/k$ in the large $N_f$ limit.
\end{abstract}

\maketitle

\section{Introduction}

While many fractionalized states of matter have been proposed, verifying their existence is a formidable task.
Not only are experimental measurements of fractional degrees of freedom difficult, but even establishing the existence of these phases in simplified lattice models can be challenging.
Numerical techniques have made a great deal of progress and now provide support for some of these states of matter.

In the context of quantum spin systems, the simplest fractionalized state with an energy gap and time-reversal symmetry
is the $\Zt$ spin liquid. Recent work described the universal spectrum of a spin system on a torus \cite{Schuler2016,Whitsitt2016}
across a transition between
a $\Zt$ spin liquid and a conventional antiferromagnetically ordered state \cite{Whitsitt2016}. Such a spectrum is a unique signature
of the transition between 
these states and goes well beyond the 4-fold topological degeneracy of the gapped $\Zt$ state that is usually examined in numerical
studies. 

In this paper, we turn our attention to critical spin liquids with an emergent photon and gapless fractionalized excitations. Commonly referred to as an `algebraic spin liquid' (ASL) or a `Dirac spin liquid', it is a critical phase of matter characterized by algebraically decaying correlators, and whose 
long-distance properties are described by an interacting conformal field theory (CFT) called 3$d$ quantum electrodynamics (QED$_3$) \cite{Rantner2001,Wen2002,Hermele2004,Karthik15,Karthik16}. 
For the kagome antiferromagnet, and also for the $J_1$-$J_2$ antiferromagnet
on the triangular lattice, there is an ongoing debate as to whether the ground state is a gapped $\Zt$ spin liquid \cite{Yan11,Jiang12,McCulloch12,ZhuWhite15,McCulloch16,Mei16}
or a U(1) Dirac spin liquid \cite{Iqbal14,Iqbal16}, and we hope our results here can serve as a useful diagnostic of numerical data.

In addition, although certain systems may not allow for an extended ASL phase, related CFTs could describe their phase transitions \cite{Grover2013,Barkeshli2013}. These `deconfined critical points' \cite{Senthil2004,Senthil2004b} require a description beyond the standard Landau-Ginzburg paradigm and are often expressed in terms of fractionalized quasiparticles interacting through a gauge field.
Our methods can be easily generalized \cite{Klebanov2012} to critical points of theories with bosonic scalars coupled to gauge fields \cite{Senthil2004,Senthil2004b},
but we will limit our attention here to the fermionic matter cases.

A close cousin of QED$_3$ can be obtained by adding an abelian Chern-Simons (CS) term to the  action. When a fermion mass is also present, the excitations of the resulting theory are no longer fermions, but instead obey anyonic statistics set by the coefficient, or `level', of the CS term. The critical `Dirac-CS' theory (with massless fermions)
has been used to describe phase transitions between fractional quantum Hall plateaus in certain limits \cite{Chen1993,Sachdev1998}
and transitions out of a chiral spin liquid state \cite{Barkeshli2013,Arun16,Lauchli16}

In this paper, we study the finite size spectrum of the QED$_3$ and Dirac-CS theories on the torus. The spectrum is a universal function of the torus circumference $L$ and modular parameter $\t$, and since numerics are often performed on this geometry, it can be used to compare with the data they generate.
The torus has the additional distinction of being the simplest topologically non-trivial manifold. A defining characteristic of topological order is the degeneracy of the groundstate when the theory is placed on a higher genus surface. 
On the torus, the pure abelian CS theory at level $k$ has $k$ ground states \cite{Witten1989,Wen1990} whose degeneracy is only split by terms
which are exponentially small in $L$.
Here, we will couple $N_f$ massless Dirac fermions to the CS theory and find a rich spectrum of low energy states with energies
which are of order $1/L$. In the limit of large $N_f$ and $k$, we will present a computation
which gives the $k$ degenerate levels in the absence of Dirac fermions and a universal spectrum with energies of order
$1/L$ in the presence of Dirac fermions. 

Proposals for ASL phases typically begin with a parton construction of the spin-1/2 Heisenberg antiferromagnet 
\eq{
H&=\sum_{\Braket{ij}}J_{ij}\v{S}_i\cdot\v{S}_j\,,
}
where $\v{S}_i$ represent the physical spin operators of the theory and $i,j$ label points on the lattice. 
Slave fermions are introduced by expressing the spin operators as 
$\v{S}_i={1\o2}f^\dag_{i\a}\bb{\s}_{\a\b}f_{i\b}$,
where $f_{i\a}$ is the fermion annihilation operator and $\bb{\s}=\(\s^x,\s^y,\s^z\)$ are the Pauli matrices. This is a faithful representation of the Hilbert space provided it is accompanied by the local constraint 
\eq{\label{eqn:gaugeConstraint}
\sum_\a f^\dag_{i\a}f_{i\a}=1.
} 
Since the physical spin $\v{S}_i$ is invariant under the transformation $f_{i\a}\rightarrow e^{i\phi_i}f_{i\a}$, the slave fermions necessarily carry an emergent gauge charge.
Replacing spins with slave fermions, decoupling the resulting quartic term, and enforcing $\Braket{f_{i\a}^\dag f_{i\a}}=1$ on average returns an ostensibly innocuous mean field Hamiltonian
$H_{\mathrm{MF}}=-\sum_{\Braket{ij}}t_{ij}f_{i\a}^\dag f_{j\a} + H.c.$ 
The mean field theory is a typical tight-binding model, but with electrons replaced by slave fermions.
However, the stability of $H_{\mathrm{MF}}$ is by no means guaranteed, and gauge fluctuations must be taken into account. This is achieved by supplementing the mean field hopping parameter with a lattice gauge connection $a_{ij}$: $t_{ij}\rightarrow t_{ij}e^{ia_{ij}}$. 
Under the renormalization group, kinetic terms for the gauge field are generated. 
Since the connection $a_{ij}$ parametrizes the phase redundancy of the $f_{i\a}$'s, it is a $2\pi$-periodic quantity, and the resulting lattice gauge theory is \emph{compact}. Determining the true fate of these theories is where numerics provide such great insight.

The mean field Hamiltonians of the models we are concerned with possess gapless Dirac cones.
In the continuum they can be expressed
\eq{\label{eqn:QED3}
S_{\mathrm{D}}[\psi,A]&=-\int d^3r\,\bpsi_{\a}i\g^\m\(\ptl_\m-i A_\m\)\psi_{\a},%+{1\o 4e^2}f_{\m\n}^2 \right\}
}
where $r=\(\t,\vx\)$ is the Euclidean spacetime coordinate, $\psi_\a$ is a two-component complex spinor  whose flavour index $\a$ is summed from 1 to $N_f$, and $A_\m$ is a U(1) gauge field that is obtained from the continuum limit of the $a_{ij}$. 
The gamma matrices are taken to be $\g^\m=\(\s^z,\s^y,-\s^x\)$, and $\bpsi_\a=i\psi_\a^\dag\s^z$.
On the the kagome lattice, the mean field ansatz with a $\pi$-flux through the kagome hexagons and zero flux through the triangular plaquettes has a particularly low energy \cite{Hastings2000,Ran2007,Hermele2008}. Its dispersion has two Dirac cones, which, accounting for spin, gives $N_f=4$.

By writing the theory in the continuum limit in the form of Eq.~\ref{eqn:QED3}, we are implicitly assuming that
monopoles (singular gauge field configurations with non-zero flux) in the lattice compact U(1) gauge theory can be neglected. 
In their absence, the usual Maxwell action can be added to the theory
\eq{\label{eqn:Maxwell}
S_\Max[A]&={1\o 4e^2}\int d^3r\, F_{\m\n}F^{\m\n}, 
&
F_{\m\n}&=\ptl_\m A_\n-\ptl_\n A_\m,
} 
resulting in the full QED$_3$ action, $S_{\mathrm{qed}}[\psi,A]=S_\Dir[\psi,A]+S_\Max[A]$.
Importantly, when $N_f$ is smaller than some critical value, these manipulations are no longer valid. $S_{\mathrm{M}}[A]$ is never an appropriate low-energy description of a lattice gauge theory with $N_f=0$: for \emph{all} values of $e^2$, monopoles will proliferate and confine the theory \cite{Polyakov1977,Polyakov1987}. 
%When $N_f=0$, the proliferation of monopoles results in the confinement of pure compact QED$_3$ for \emph{all} values of $e^2$ \cite{Polyakov1977,Polyakov1987}, and $S_{\mathrm{M}}[A]$ is never an appropriate low-energy description;
In the confined phase, the slave fermions cease to be true excitations, and remain bound within the physical spins $\v{S}_i$.
However, matter content suppresses the fluctuations of the gauge field. 
For $N_f$ large enough, monopoles are irrelevant operators, \cite{Borokhov2002,Hermele2004,Pufu14,Pufu15} and $S_{\mathrm{qed}}[\psi,A]$ is a stable fixed point of the lattice theory \cite{Hermele2004}. 
In this limit, QED$_3$ is believed to flow to a non-trivial CFT in the infrared, and this has been shown perturbatively to all orders in $1/N_f$ \cite{Appelquist1981,Jackiw1981,Templeton1981,Templeton1981b}. 
The critical theory is obtained by na\"{i}vely taking the limit $e^2\rightarrow\infty$, and, for this reason, the Maxwell term will be largely ignored in what follows.

The Dirac fermions $\psi_\a$ represent particle or hole-like fluctuations about the Fermi level.
Consequently, any single-particle state violates the local gauge constraint in Eq.~\ref{eqn:gaugeConstraint} and is prohibited. 
Since fluctuations in $A_\m$ are suppressed at $N_f=\infty$, we might expect this neutrality to be the only signature of the gauge field
in the large $N_f$ limit, and so the spectrum on the torus is given by the charge neutral multi-particle states of the free field theory.
It is important to note that all of these multi-particle states are built out of single fermions $\psi_\a$ which obey {\it anti-periodic\/}
boundary conditions around the torus: such boundary conditions (or equivalently, a background gauge flux of $\pi$ and periodic
boundary conditions for the fermions) minimize the ground state energy, as we show in Appendix~\ref{app:freeFermion}. 
Some of these energy levels are given in Table~\ref{tab:k0FreeFermEnergies}. 

Even among the charge neutral multiparticle states, there are certain states of the free field theory which are strongly renormalized even
at $N_f = \infty$. These are the SU($N_f$) singlet states which couple to the $A_\mu$ gauge field. Computation of these
renormalizations is one of the main purposes of the present paper. We show that the energies of these states are instead given by the zeros of the gauge field effective action. A similar conclusion was reached
in Ref.~\onlinecite{Whitsitt2016} for the O($N$) model, where the O($N$) singlet levels were given by the zeros of the effective action
of a Lagrange multiplier. 

{\def\arraystretch{0.65}%  1
\begin{table}
\centering
\begin{tabular}{| c | c || c | c || c | c |}
%\multicolumn{12}{c}{External momentum $\bar{\vq}$}\\\hline
\hline
\multicolumn{2}{|c||}{$\bar{\vq}=(0,0)$} & \multicolumn{2}{c||}{$\bar{\vq}=(1,0)$} & \multicolumn{2}{c|}{$\bar{\vq}=(1,1)$} \\ \hline
 $\bar{\w}_\g$ & $d_\g$ &$\bar{\w}_\g$ &$d_\g$ & $\bar{\w}_\g$ & $d_\g$ \\\hline\hline
 0.584130 &{2}& & &  & \\
 & & 1.437980  & 1 & &\\
 & & & & 1.682078 & 1 \\
 & & & &   1.739074 & 1\\
 & &1.976292 & 1  & & \\
 2.311525 & 2 & & & & \\
 & & & &2.527606 & 1\\
  & & 2.658092  &1  & & \\
  & & & &   2.813224&1 \\
 & & 3.156341  &1 & & \\
    & & 3.407832 &1 &  & \\
& & & &  3.517617 & 1 \\
 &  & & &  3.626671&1 \\
 & & 3.814432  &1  & & \\
 3.855225 & 2 &  & & & \\
 & & 4.092996   &1 &  & \\
 & & & &   4.259784 & 1\\
& & & &  4.330137 & 1\\
  & &4.425387 & 1& &  \\
 & & &  & 4.523167 & 1\\
 4.586816 & 2 & & & & \\
 & & & &   4.657172 & 1 \\
 & &  4.685590& 1&  & \\\hline
  %\hline
% & &  & &  5.036796 & 4 & & & &  & &  \\
% 5.099020 & 8 & & & 5.099020 & 2 &  & && & & \\
%  & &  & &  5.464985 & 4 & & & & & &  \\
% & & 5.489309 & 2 & & & & & & & & \\
\end{tabular}
\caption{Photon modes in QED$_3$ (CS level $k=0$) on a square torus of size $L$. 
Frequencies are shown for $\vq=0$, $\vq_1=2\pi(1,0)/L$, and $\vq_2=2\pi(1,1)/L$. 
The 1st, 3rd, and 5th columns list the frequencies, $\w_\g$, while the column  immediately to the right provides the degeneracy, $d_\g$. 
The actual photon energy levels are given by these frequencies as well as integer multiples.
($\bar{\vq}=L\vq/2\pi$, $\bar{E}=LE/2\pi$.)}\label{tab:k0Energies}
\end{table}}

In Table~\ref{tab:k0Energies}, we list some of the lowest frequency modes of the photon in QED$_3$ on a square torus, obtained in the large $N_f$ computation
just described. 
Because the theory on the torus is translationally invariant, we can distinguish states by their total external momentum. For each momentum considered, the left-most column gives the photon frequency with its degeneracy is shown on the right. 
By including multi-photon states, the actual energy levels of the photon are shown in Table~\ref{tab:k0EnergyLevels} for the same set of momenta. 
%free-fermion energy levels, while the third column shows the shifted energies of the SU($N_f$) 
%singlet states obtained from the photon propagator. To the right of each, the degeneracy of the level is provided. 
The origin of the photon shift will be apparent when we  find the free energy in Sec.~\ref{sec:freeEnCSD} and explicitly calculate the energy levels in Sec.~\ref{sec:Spectrum}.

A similar story applies to the Dirac-CS theory with finite CS coupling $k$:
\eq{\label{eqn:CSaction}
S_{\mathrm{CS}}[A]&
={ik\o4\pi}\int d^3r\,\ep^{\m\n\r}A_\m\ptl_\n A_\r. 
} 
The addition of this term gives the photon a mass and attaches flux to the Dirac fermions so that they become anyons with statistical angle $\theta=2\pi(1-1/k)$.
The Dirac-CS theory applies to the chiral spin liquid which spontaneously breaks time reversal, generating a Chern-Simons term at level $k=2$ \cite{Kalmeyer1987}.
Similarly, a CS term with odd level can be used to impose anyonic statistics on the quasiparticles of a fractional quantum Hall fluid. 
The Dirac-CS CFT we consider can describe the continuous transitions into and between such topological phases \cite{Barkeshli2013,Arun16,Lauchli16}. It is given by $S_{\mathrm{DCS}}[\psi,A]=S_{\mathrm{D}}[\psi,A]+S_{\mathrm{CS}}[A]$ (after taking $e^2\rightarrow\infty$). 
As $k$ becomes very large, the anyons become more fermion-like, making an expansion in $2\pi/k$ possible at large $N_f$ \cite{Chen1993,Sachdev1998}. 

Once again, keeping $\lam=N_f/k$ fixed, the critical 
Dirac-CS theory is both stable and tractable in the large-$N_f$ limit. The qualitative features of the spectrum are very similar to QED$_3$. 
Again $\psi_\a$ is not a gauge invariant quantity and cannot exist by itself in the spectrum. 
The Gauss law mandates that it be accompanied by $k$ units of flux. In the large-$k$ limit, these states have very high energies and can be neglected: only charge-neutral excitations need be considered.
Likewise, the energy levels of the SU($N_f$) singlet states coupling to the gauge field are strongly renormalized even 
at large $N_f$, while the mixed-flavor two-particle excitations behave as free particles.
As $k/N_f$ becomes large, the Chern-Simons term will dominate and the topological degeneracy which was lost upon coupling to matter will reassert itself. 
The photon modes of the zero external momentum sector are shown in Table~\ref{tab:kfiniteEnergies} for several values of $\lam$.

We will calculate the energy spectrum using a path integral approach similar to that of Ref.~\onlinecite{Klebanov2012}. 
In order to ensure that the gauge redundancy is fully accounted for, it is useful to first calculate the free energy. 
This is done in Sec.~\ref{sec:pathInt}, starting with two exactly solvable theories, pure Chern-Simons and Maxwell-Chern-Simons, before moving on to QED$_3$ and the Dirac-CS theory in the large-$N_f$ limit. The structure of the free energy will allow us
to identify the multi-fermion states, along with their bound states which appear in the photon contribution.
In Sec.~\ref{sec:Spectrum} we determine the energy levels and we conclude in Sec.~\ref{sec:conclusion}.

\section{Path integral and free energy}\label{sec:pathInt}

To understand the spectrum of the large-$N_f$ QED$_3$ and Dirac-CS theory, we evaluate its path integral \cite{Klebanov2012}. The path integral is
\eq{
Z&={1\o\text{Vol}(G)}\int DA\, D\psi\,e^{-S[A,\psi]}
}
where $\text{Vol}(G)$ is the volume of the gauge group. 
For simplicity, we work on the square torus: the modular parameter $\t=i$ and the $x$- and $y$-cycles are equal in length: $x\sim x+L$, $y\sim y+L$. Eventually, we will specify to the zero-temperature limit, $1/T=\b\rightarrow\infty$, but for now we leave $\b$ finite.

The gauge field $A$ can be split into zero and finite momentum pieces, 
\eq{\label{eqn:gaugeDecomp}
A_\m&=a_\m+A'_\m,
&
A'_\m&={1\o \sqrt{\b L^2}}\sum_p^\prime A_\m(p)e^{ipr},
}
where $p$ sums over $p_\m=2\pi n_\m/L_\m$, $L_\m=(\b,L,L)$ where $n_\m\in\Z$. The prime on the summation indicates that the $n_\m=(0,0,0)$ mode is not included.
The measure of integration is therefore $DA=Da\, DA'$.
Unlike on $\mathds{R}^3$, the zero modes $a$ are not pure gauge configurations. Instead, the gauge transformation which shifts $a$,
\eq{
U&=\exp\[{2\pi i}\sum_\m {n_\m r_\m \o L_\m}\],
}
is only well-defined provided $n_\m\in\Z$.
Under the action of $U$, the zero modes transform as $a_\m\rightarrow a_\m+2\pi n_\m/L_\m$, and so they are
periodic variables and should be integrated only over the intervals $[0,2\pi/L_\m)$. 
Including a Jacobian factor of $\sqrt{\b L^2}$ for each component, we have
\eq{\label{eqn:zeroModeInt}
\int Da&=\(\b L^2\)^{3/2}\int_0^{2\pi/\b} d a_0 \int_0^{2\pi/L}d^2\bb{a}.
}

The spatially varying portion of the gauge field can be decomposed further into $A'=B+d\phi$ where $\phi$ parametrizes the gauge transformations of $A'$, and $B$ may be viewed as the gauge-fixed representative of $A'$. 
Naturally, gauge invariance implies that the action is independent of $\phi$: $S[\psi,A]=S[\psi,a+B]$.
Here, we work in the Lorentz gauge, $\ptl^\m B_\m=0$.
The full measure of integration is then 
\eq{
DA&=Da\,DB\,D(d\phi).
}
We begin by expressing $D(d\phi)$ directly in terms of the phases $\phi$. 
They can be related through the distance function $\D(\w,\w+\d\w)=\(\int\abs{\d\w}^2\)^{1/2}$:
\eq{
\D\(\phi,\phi+\d\phi\)&=\(\int\abs{\d\phi}^2\)^{1/2}
\nt
\D(d\phi,d\phi+d\d \phi)&=\(\int\abs{d\d\phi}^2\)^{1/2}=\(\int \d\phi\(-\nabla^2\)\d\phi\)^{1/2}.
}
Changing variables, the measure becomes
\eq{\label{eqn:dphiMeasure}
D(d\phi)&=D'\phi\sqrt{\text{det}'\(-\nabla^2\)}
}
where the primes indicate that constant configurations of $\phi$ are not included and that the zero eigenvalue of the Laplacian is omitted. 
This functional determinant is the familiar Faddeev-Popov (FP) contribution to the path integral.
As expected for abelian gauge theories, both of these factors are independent of the gauge field $B$.

The volume of the gauge group can be divided in a similar fashion
\eq{
\text{Vol}(G)&=\text{Vol}(H)\int D'\phi,
}
where $H$ is the group of constant gauge transformations. $\int D'\phi$ will cancel the identical factor present in the numerator from the gauge field measure in Eq.~\ref{eqn:dphiMeasure}, and 
$\text{Vol}(H)$ can be determined using the distance function defined above. A constant gauge transform has $\phi=c$, a constant, where $c\in\[0,2\pi\)$. We find
\eq{
\text{Vol}(H)&=\int_0^{2\pi}dc\,{\D(c,c+\d c)\o\d c}=\int_0^{2\pi}dc\,{\d c\o\d c}\(\int 1\)^{1/2}
\nt
&=2\pi\sqrt{\text{Vol}\(\v{T}^2\times\v{S}^1\)}=2\pi\sqrt{\b L^2}.
}

Putting these facts together, we are left with
\eq{\label{eqn:partFunFull}
Z&={\b L^2\o2\pi}\sqrt{\text{det}'\(-\nabla^2\)}\int d^3a\,DB\, D\psi\,e^{-S[a,B,\psi]}.
}
In the following two sections, we calculate the free energies and partition functions of the pure Chern-Simons and the Maxwell-Chern-Simons theories. These serve as simple examples (and verifications) of the normalization and regularization procedure, before we move on to the third section and primary purpose of this paper, large-$N_f$ QED$_3$ and Dirac-Chern-Simons. 

\subsection{Pure Chern-Simons theory}\label{sec:Fcs}

It is well-known that pure abelian Chern-Simons theory should have $Z_{\mathrm{CS}}=k$ \cite{Witten1989}. 
Since the action in Eq.~\ref{eqn:CSaction} only has linear time derivatives, the Hamiltonian vanishes and it may at first be surprising that $Z_{\mathrm{CS}}$ is not simply unity: $\braket{0|0}=1$. 
One way to understand this is through canonical quantization. 
The observable operators of the theory are the two Wilson loops winding around either cycle of the torus. Their commutations relations are determined by the Chern-Simons term, and at level $k$, it can be shown that the resulting representation requires at least a $k$-dimensional Hilbert space (see e.g. \cite{Poly1990}). The partition function is therefore $Z_{\mathrm{CS}}=\sum_{n=1}^k\Braket{n|n}=k$. 
Within the general framework of topological field theories, the partition function on the torus should evaluate to the dimension of the corresponding quantum mechanical Hilbert space.

The pure CS partition function is
\eq{
Z_{\CS}&={\b L^2\o2\pi}\sqrt{\text{det}'\(-\nabla^2\)}\int da\,DB\,e^{-S_{\CS}[B]}.
}
We write the Chern-Simons action in momentum space as $S_{\CS}[B]={1\o2}\sum_q B_\m(-q)\Pi_{\CS}^{\m\n}(q)B_\n(q)$ where 
\eq{\label{eqn:PiCSpure}
\Pi_{\CS}^{\m\n}(q)&={ik\o2\pi}\ep^{\m\n\r}q_\r,
}
with $q_\m=2\pi n_\m/L_\m$, $n_\m\in\Z$. Performing the Gaussian integral, we find
\eq{
Z_{\CS}&={\b L^2\o2\pi}\sqrt{\text{det}'\(-\nabla^2\)}\sqrt{\text{det}'\(2\pi\o\Pi_{\CS}^{\m\n}\)}\int da.
}
It is simpler to work with the free energy and then return to the partition function at the end of the calculation:
\eq{\label{eqn:CSFtot}
F_{\CS}&=-{1\o\b}\log Z_{\CS}=
F_{a}+F_\pi+F_{\FP}-{1\o\b}\log\[ \b L^2\o 2\pi\].
}

We proceed to treat each contribution individually. The integral over the zero modes gives
\eq{\label{eqn:CSFzm}
F_{a}&=-{1\o\b}\log\[\int da\]=-{1\o\b}\log\[ (2\pi)^3\o \b L^2\].
}
This cancels the volume-dependent constant in the free energy, leaving $F_{\CS}=-{1\o\b}\log(2\pi)^2+F_\pi+F_{\FP}$.
The FP determinant's contribution is 
\eq{\label{eqn:CSFfp}
F_{\FP}&=-{1\o\b}\log\sqrt{\text{det}'\(-\nabla^2\)}=-{1\o2\b}\sum_q'\log q^2
}
where $q_\m=2\pi n_\m/L_\m$, $n_\m\in \Z$. As will be the convention throughout this paper, the prime on the summation indicates that the zero momentum mode ($n_\m=(0,0,0)$) is omitted.
Finally, the piece from the Gaussian integral is
\eq{
F_\pi&={1\o2\b}\log\text{det}'\[ \Pi^{\m\n}_{CS}\o2\pi\].
}
For each momentum $q_\m$, the Chern-Simons kernel has three eigenvalues, 0 and $\pm ik\abs{q}/2\pi$, but only the non-zero values should be included. 
In fact, it is easy to verify that the eigenvector corresponding to the 0 eigenvalue is proportional to $q_\m$ and consequently arises from the pure gauge configurations $\sim \ptl_\m\phi$ which have already been accounted for.
Therefore,
\eq{
F_\pi&={1\o2\b}\sum_q'\log\[ {1\o 4\pi^2}{k^2\o4\pi^2} q^2\].
}
Using the zeta-function regularization identity $\sum_p'=-1$, we have
\eq{\label{eqn:CSFg}
F_\pi={1\o2\b}\sum_q'\log q^2-{1\o\b}\log\({k\o 4\pi^2}\).
}
The momentum sum in $F_\pi$ cancels exactly with the sum in $F_{\FP}$. This is a direct consequence of the fact that the CS theory has no finite energy states and, notably, is only apparent when the Faddeev-Popov and gauge kernel determinants are considered together. 
All together, the total free energy is
\eq{
F_{\CS}&=-{1\o\b}\log k,
}
which gives $Z_{\CS}=k$ as claimed.

\subsection{Maxwell-Chern-Simons theory}

It is also useful to understand how the topological degeneracy emerges in the presence of finite-energy modes. This is easily accomplished by adding a Maxwell term: 
\eq{
S_{\mathrm{MCS}}[A]&=S_\Max[A]+S_\CS[A],
}
where $S_\Max[A]$ is given in Eq.~\ref{eqn:Maxwell}. 
The procedure for calculating the free energy is identical to the pure CS case except that the gauge kernel is now 
\eq{
\Pi_\mathrm{MCS}(q)&={q^2\o e^2}\(\d^{\m\n}-{q^\m q^\n\o q^2}\)+{ik\o2\pi}\ep^{\m\n\r}q_\r.
}
As above, this matrix has one vanishing eigenvalue in the pure gauge direction and two non-trivial ones in orthogonal directions:
\eq{
{q^2\o e^2}\pm{ik\o 2\pi}\abs{q}.
}

Performing the functional integral and taking the logarithm, we find
\eq{
F_\pi &={1\o2\b}\sum_q'\log\[{q^4\o e^4}+{k^2q^2\o4\pi^2}\].
}
As in the pure CS case, the FP determinant cancels a factor of $q^2$ from $F_\pi$. Now, however, this does not completely remove the momentum dependence of the sum. The total free energy is 
\eq{\label{eqn:FmcsFreq}
F_{\mathrm{MCS}}&=-{1\o\b}\log 4\pi^2+F_\pi+F_\FP={1\o\b}\log \( e^2\o2\pi\)+{1\o2\b}\sum_{n,\vq}'\log\[ \ep_n^2+\vq^2+{e^4k^2\o4\pi^2}\],
}
where we've written $q^\m=\(\ep_n,\vq\)$ with $\ep_n=2\pi n/\b$, $n\in\Z$. 
Analytically continuing to real time, $\ep_n\rightarrow -i\w$, the argument of the logarithm is $\w^2-\g_\vq^2$ where $\g_\vq=\sqrt{\vq^2+\(e^2k/2\pi\)^2}$. 
We recognize the $\g_\vq$'s as the frequencies of a set of harmonic oscillators. As in the previous section, this is only manifest when the sum $F_\pi+F_\FP$ is considered: by itself, $F_\pi$ seems to imply the existence of an extra set of oscillators whose frequencies  are $\tilde{\g}_\vq=\abs{\vq}$.

The presence of the oscillators is even clearer upon performing the (imaginary) frequency sum. Adding and subtracting the zero mode, we are left to evaluate an infinite sum
\eq{\label{eqn:FmcsSum}
F_{\mathrm{MCS}}&=-{1\o\b}\log\(2\pi\g_0\o e^2\)+{1\o2\b}\sum_{n,\vq}\log\[n^2+\(\b\g_\vq\o2\pi\)^2\].
} 
By using the known analytic properties of the zeta function for complex $s$, we can assign a value to the otherwise obviously diverging sum. 
For the logarithm, this representation results in the identification
\eq{\label{eqn:EpsteinZeta}
\sum_{n}\log\[n^2+\(\b\g_\vq\o2\pi\)^2\]=-\lim_{s\rightarrow0}{d\o ds}\sum_n\[n^2+\(\b\g_\vq\o2\pi\)^2\]^{-s}=-\lim_{s\rightarrow0}{d\o ds}\zeta_\mathcal{E}\(s;\(\b\g_\vq\o2\pi\)^2\)
}
where $\zeta_\mathcal{E}(s;a^2)$ is the Epstein zeta function. After some standard manipulations (given in Appendix~\ref{app:MaxCS}), we arrive at the expression 
\eq{
F_{\mathrm{MCS}}&=-{1\o\b}\log k-{1\o\b}\sum_\vq\log\[ e^{-\b\g_\vq/2}\o 1-e^{-\b\g_\vq}\].
}
Re-exponentiating, we find 
\eq{
Z_{\mathrm{MCS}}&=k\prod_\vq Z_\vq,
&
Z_\vq&={e^{-\b\g_\vq/2}\o1-e^{-\b\g_\vq}}=e^{-\b\g_\vq/2}\sum_{n=0}^\infty e^{-\b n\g_\vq}.
}
As observed, the partition function is a product over an infinite stack of harmonic oscillators with frequencies $\g_\vq$.
The topological degeneracy enters through the factor of $k$ multiplying $Z_\CS$: there are $k$ identical sets of oscillators. 
We note that in the limit $e^2\rightarrow\infty$, the barrier to the first excited state becomes infinitely large, effectively projecting onto the lowest Landau level. Ignoring some constants, we arrive back at the pure Chern-Simons described above.

\subsection{QED$_3$ and Dirac-Chern-Simons theory}\label{sec:freeEnCSD}

When we couple the gauge field to fermions, the partition function is no longer exactly solvable. 
Nonetheless, when the number of fermion flavours, $N_f$, is large, a saddle-point approximation is valid and allows a systematic expansion in $1/N_f$. 
As discussed in the introduction, the QED$_3$ and Dirac-CS fixed points are obtained in the limit $e^2\rightarrow\infty$, and so we will not explicitly include the Maxwell action $S_\Max[A]$ in our calculations. 
In order to avoid the parity anomaly \cite{Redlich1984,Redlich1984b}, we take $N_f$ to be even in all that follows.
The partition function is given in Eq.~\ref{eqn:partFunFull} with action 
\eq{
S_{\mathrm{DCS}}[\psi,A]&=S_\Dir[A,\psi]+ S_\CS[A].
}
where $S_\Dir[A,\psi]$ and $S_\CS[A]$ are given in Eqs.~\ref{eqn:QED3} and \ref{eqn:CSaction} respectively. 
The Chern-Simons level $k$ is assumed to be of the same order as $N_f$.
We begin by integrating out the fermions,
\eq{
Z&={\b L^2\o2\pi}\sqrt{\text{det}'\(-\nabla^2\)}\int da\,DB\,\exp\(-S_{\CS}[B]+N_f\log \det i \sd{D}\),
}
where $\sd{D}=\s^\m\(\ptl_\m-ia_\m-iB_\m\)$. We subsequently expand the determinant in terms of $B$:
\eq{\label{eqn:fermiDetExp}
\log\det(i\sd{D})&=\tr\log\(i\sd{\ptl}+\sd{a}\)+\tr\({1\o i\sd{\ptl}+\sd{a}}\,\sd{B}\)-{1\o 2}\tr\( {1\o i\sd{\ptl}+\sd{a}}\,\sd{B}\,{1\o i\sd{\ptl}+\sd{a}}\,\sd{B}\)+\cdots
}
By rescaling $B\rightarrow B/\sqrt{N_f}$, the subleading behaviour of the linear and quadratic terms, as well as the Chern-Simons action, is clear. 

On the plane, the saddle-point value of $A$ vanishes by symmetry and gauge invariance. However, since $A\rightarrow A+c$ for constant $c$ is no longer a gauge transformation on the torus, the zero modes are distinct and could conceivably have a non-zero expectation value: $\Braket{a}=\bar{a}\neq0$. 
In fact, neither the pure CS nor Maxwell-CS actions depended on $a_\m$. 
The matter lifts this degeneracy by creating an effective potential for the $a$'s, and 
$\bar{a}$ can be determined by minimizing the free fermion functional determinant
\eq{
F_0(a)&=-\tr\log\(i\sd{\ptl}+\sd{a}\)=-\sum_p\log\(p+a\)^2.
}
The summation above is over spacetime momenta $p_\m={2\pi \( n_\m+1/2\)/L_\m}$, $n_\m\in \Z$ as is appropriate for our choice of fermions with antiperiodic boundary conditions. 
This calculation is performed in Appendix~\ref{app:freeFermion} where it is shown that the saddle-point value of the gauge field is $\bar{a}_\m=0$: this is closely linked to the choice of anti-periodic boundary conditions for the fermions, which we have established also minimize
the total energy. 

The linear term in $B$ in Eq.~\ref{eqn:fermiDetExp} vanishes, so that the subleading term in the determinant expansion is
\eq{
S_{f}[B]&={N_f\o2}\tr\({1\o i \sd{\ptl}}\sd{B}{1\o i\sd{\ptl}}\sd{B}\)
={N_f\o2}\sum_q B_\m(-q)\Pi_f^{\m\n}(q)B_\n(q)
}
where 
\eq{
\Pi^{\m\n}_f(q)&={2\o\b L^2}\sum_p{ p^\m\(p^\n+q^\n\)+\(p^\m+q^\m\)p^\n-\d^{\m\n}p\cdot \(p+q\) \o p^2\(p+q\)^2}.
}
%This is actually just the current-current correlator, $\Braket{J^\m(q)J^\n(-q)}$ shown in Fig.~\ref{fig:JJcorr}.
On the plane, this expression evaluates to \cite{Kaul2008}
\eq{\label{eqn:PiInf}
\Pi_{\infty}^{\m\n}={\abs{q}\o16}\(\d^{\m\n}-{q^\m q^\n\o q^2}\).
}
On the torus, a simple analytic formula is no longer available and $\Pi_f$ must be calculated numerically. Expressions for  the components of $\Pi^{\m\n}_f$ on the symmetric torus are given in Appendix~\ref{app:Pif}.
%\begin{figure}
%\centering
%\includegraphics[scale=0.35]{JJcorr.pdf}
%\caption{Current-current correlator}\label{fig:JJcorr}
%\end{figure}

Since $k\sim\O(N_f)$, the CS term will contribute at the same order as $\Pi_f$. Rescaling Eq.~\ref{eqn:PiCSpure} to bring out an overall factor of $N_f$, we write the momentum space kernel of the Chern-Simons term as
\eq{
\Pi_{\CS}^{\m\n}(q)& ={i\o2\pi\lam}\ep^{\m\n\r}q_\r,
&
\lam&={N_f\o k}\,.
}
All together, the full effective potential is 
\eq{\label{eqn:Seff}
S_{\text{eff}}[B]&={N_f\o2}\sum_q B_\m(-q)\Pi^{\m\n}(q)B_\n(q),
&
\Pi^{\m\n}(q)&=\Pi_{\CS}^{\m\n}(q)+\Pi^{\m\n}_f(q),
}
and the large-$N_f$ partition function is
\eq{\label{eqn:pathIntLargeN}
Z&\cong{\b L^2\o2\pi}\sqrt{\text{det}'\(-\nabla^2\)}\,e^{-\b N_fF_0(\bar{a})}\int DB\,\exp\[-{1\o2}\sum_q B_\m(-q)\Pi^{\m\n}(q)B_\n(q)\]
\nt
&={\b L^2\o2\pi}\sqrt{\text{det}'\(-\nabla^2\)}\,e^{-\b N_fF_0(\bar{a})}\sqrt{\text{det}'\(2\pi\o\Pi^{\m\n}\)}.
}
The corresponding free energy is
\eq{\label{eqn:Ftot}
F&=-{1\o\b}\log Z\cong N_fF_0+F_G-{1\o\b}\log\[ \b L^2\o 2\pi\]
}
where the full gauge field contribution is
\eq{
F_\Gg&=F_{\FP}+F_\pi
\nt
F_\pi&=-{1\o\b}\log\sqrt{\text{det}'\(2\pi\o \Pi^{\m\n}\)}
&
F_{\FP}&=-{1\o\b}\log\sqrt{\text{det}'\(-\nabla^2\)}.
}

%\subsection{Dirac-CS theory}
\subsubsection{Zero external momentum, $\vq=0$}
We begin by considering the zero (spatial) momentum portion of the free energy. Denoting the Euclidean spacetime momenta $q^\m=\(\ep,\vq\)$, we set $\vq=0$.
In this case, only $\Pi^{ij}(\ep,0)\neq0$, for $i,j=x,y$:
\eq{
\Pi^{ij}(\ep,0)&=
\begin{pmatrix}
\Pi_f^{xx} & \ep/2\pi\lam \\
-\ep/2\pi\lam & \Pi^{yy}_f
\end{pmatrix}.
} 
Expressions for $\Pi^{xx}_f$ and $\Pi^{yy}_f$ are given in Eqs.~\ref{eqn:Pixx} and \ref{eqn:Pixy} of Appendix~\ref{app:Pif}.
Taking the determinant, the free energy is
\eq{
F^{\vq=0}_\pi&={1\o\b}\log 2\pi + {1\o 2\b}\sum_n'\log\[ \Pi_f^{xx}\(\ep_n,0\)^2+{\ep_n^2\o 4\pi^2\lam^2}\]
}
where  $\ep_n=2\pi n/\b$, $n\in\Z/\{0\}$, and the symmetry of the torus has been used to set $\Pi_f^{xx}=\Pi_f^{yy}$. The FP piece is
\eq{
F_{\FP}^{\vq=0}&=-{1\o2\b}\sum'_n\log\ep_n^2\,.
}
Adding the two and taking the zero temperature limit, $\b\rightarrow\infty$, the total gauge contribution is 
\eq{
 F^{\vq=0}_\Gg&=
 {1\o 2}\int {d\ep\o2\pi}\log\[ \(\Pi_f^{xx}\o\ep\)^2+{1\o 4\pi^2\lam^2}\].
}
For large $\ep$, the integral does not converge. Instead, $\Pi_f^{xx}$ approaches its infinite volume limit in Eq.~\ref{eqn:PiInf}:
\eq{
\(\Pi^{xx}_f\o\ep\)^2+{1\o4\pi^2\lam^2}\rightarrow \(1\o16\)^2+{1\o4\pi^2\lam^2}\,.
}
This is not a problem since an integral over a constant vanishes in the zeta regularization scheme. Adding and subtracting the large frequency limit, the free energy is a finite function
\eq{\label{eqn:Fq0}
 F^{\vq=0}_\Gg&={1\o 2}\int {d\ep\o2\pi}
\Bigg\{\log\[ \(\Pi_f^{xx}\o\ep\)^2+{1\o 4\pi^2\lam^2}\]-\log\[\(1\o16\)^2+{1\o4\pi^2\lam^2}\]\Bigg\}.
}

\subsubsection{Finite external momentum, $\vq\neq0$}

For the finite momentum piece, we begin by restricting the polarization matrix $\Pi^{\m\n}(\ep,\vq)$ to the physical subspace. %its physical directions.
As required by gauge invariance,
it has a vanishing eigenvalue along the $q^\m=(\ep,\vq)$ direction: $q_\m\Pi^{\m\n}=0$. To determine the remaining two modes, we project onto the orthogonal directions
\eq{
v_T&={1\o\abs{\vq}}\begin{pmatrix} 0 \\ q_y \\ -q_x \end{pmatrix},
&
v_L&={1\o\abs{\vq}\sqrt{\ep^2+\vq^2}}\begin{pmatrix} -\vq^2 \\ \ep q_x \\ \ep q_y \end{pmatrix},
}
and, after some simplifying, arrive at
\eq{
\Pi_{\text{proj}}&={1\o\vq^2}
\begin{pmatrix}
\(\ep^2+\vq^2\)\Pi^{00} & \sqrt{\ep^2+\vq^2}\(q_y\Pi^{0x}-q_x\Pi^{0y}\) \\
\sqrt{\ep^2+\vq^2}\(q_y\Pi^{0x}-q_x\Pi^{0y}\) & \vq^2\(\Pi^{xx}+\Pi^{yy}\)-\ep^2\Pi^{00}
\end{pmatrix}.
}
Taking the determinant, the contribution to the free energy is
\eq{\label{eqn:FpiQn0}
F^{\vq\neq0}_\pi&=-{1\o\b}\log\sqrt{\text{det}'\(2\pi\o \Pi^{\m\n}\)}=-{1\o \b}\sum'_{\ep,\vq}\log2\pi+{1\o 2\b}\sum'_{\ep,\vq}\log\Pi^{\m\n}
\nt
&={1\o 2}\int{d\ep\o2\pi}\sum'_{\vq}\log\Bigg\{{\(\ep^2+\vq^2\)\o\vq^2}\[ {\Pi^{00}}\(\Pi^{xx}+\Pi^{yy}-{\ep^2\o\vq^2}\Pi^{00}\)-{1\o \vq^2}\(q_y\Pi^{0x}-q_x\Pi^{0y}\)^2\]\Bigg\}
}
where the ${1\o\b}\log2\pi$ term has vanished in the zero temperature limit.
The Faddeev-Popov portion of the free energy,
\eq{
F^{\vq\neq0}_{\FP}&%=-{1\o \b}\log\sqrt{\text{det}'\(-\nabla^2\)}%=-{1\o 2\b}\sum_{\ep,\vq}\log\[ \ep^2+\vq^2\]
=-{1\o2}\int{d\ep\o2\pi}\sum_\vq\log\(\ep^2+\vq^2\),
}
perfectly cancels the $\ep^2+\vq^2$ prefactor inside the logarithm in Eq.~\ref{eqn:FpiQn0}.
Had it not been included, we may have erroneously assumed the existence of a state with energy $E=\abs{\vq}$ as there is on the plane when $k=0$. 

As $\ep^2+\vq^2$ becomes large, $\Pi^{\m\n}$ approaches its infinite volume limit (Eq.~\ref{eqn:PiInf}) like in the $\vq=0$ case. Here as well, the summand becomes a constant which vanishes in our regularization procedure.
Putting this together, we have
\eq{\label{eqn:Fqn0}
F_\Gg^{\vq\neq0}&={1\o2}\int{d\ep\o2\pi}\sum'_{\vq}\Bigg\{
\log\[ {\Pi^{00}\o\vq^2}\(\Pi^{xx}+\Pi^{yy}-{\ep^2\o\vq^2}\Pi^{00}\)-{1\o \vq^4}\(q_y\Pi^{0x}-q_x\Pi^{0y}\)^2\]
\nt
&\quad-\log\[\(1\o16\)^2+{1\o4\pi^2\lam^2}\]\Bigg\}.
}
The total contribution of the gauge field to the free energy is given by the sum of this expression with $F_G^{\vq=0}$ in Eq.~\ref{eqn:Fq0}.

\section{Spectrum}\label{sec:Spectrum} 

In this section we explicitly calculate the universal spectrum on the finite torus using the path integral expansion we just derived.

As the photon is the only element of the theory which differs from the free theory of $N_f$ Dirac fermions, it is not surprising that the free theory spectrum can account for most of the states.
The free Hamiltonian is
\eq{
\mathcal{H}_\Dir&=-i\int d^2\vx\,\psi^\dag_\a(\vx)\s_i\ptl_i\psi_\a(\vx),
}
and can be diagonalized by first going to Fourier space,
\eq{
\psi_\a(\vx)&={1\o L^2}\sum_\vp e^{i\vq\cdot\vx}\begin{pmatrix}c_{1\a}(\vp)\\ c_{2\a}(\vp)\end{pmatrix},
&
\vp&={2\pi\o L}\( n_x+{1\o2},n_y+{1\o2}\),\quad n_{x,y}\in \Z,
}
and then changing basis to $\chi_{\pm\a}(\vp)$:
\eq{
\begin{pmatrix}c_{1\a}(\vp)\\ c_{2\a}(\vp)\end{pmatrix}
&=
{1\o\sqrt{2}}
\begin{pmatrix}
1 & 1 \\ {P/\abs{\vp}} & - {P/ \abs{\vp}}
\end{pmatrix}
\begin{pmatrix}
\chi_{+\a}(\vp) \\ \chi_{-\a}(\vp)
\end{pmatrix},
}
where $P=p_x+i p_y$, $\abs{\vp}=\sqrt{p_x^2+p_y^2}$. In this basis, the Hamiltonian is
\eq{\label{eqn:HamDiag}
\mathcal{H}_\Dir=\sum_\vp\abs{\vp}\[\chi^\dag_{+\a}(\vp)\chi_{+\a}(\vp)-\chi_{-\a}^\dag(\vp)\chi_{-\a}(\vp)\].
}
We identify the vacuum as the state having all negative energy modes filled: $\chi_{+\a}(\vp)\ket{0}=\chi^\dag_{-\a}(\vp)\ket{0}=0$. 
Consequently, $\chi_{+\a}^\dag(\vp)$ is a particle creation operator carrying momentum $\vp$, and $\chi_{-\a}(\vp)$ is a hole creation operator carrying momentum $-\vp$. 
Note that all the fermionic momenta correspond to anti-periodic boundary conditions around the
torus, because these minimize the ground state energy, as shown in Appendix~\ref{app:freeFermion}.

To determine the excitations relevant to QED$_3$ and the Dirac-CS theory, we recall that once the theory is gauged, neither $\chi_{+\a}(\vp)$ nor $\chi_{-\a}(\vp)$ is gauge invariant, and all single-particle states are prohibited. Similarly, only charge-neutral two-particle states are allowed. We therefore expect the lowest fermion-like energy states to be of the form
\eq{\label{eqn:2particle}
&\chi_{+\a}^\dag(\vp+\vq)\chi_{-\b}(\vp)\Ket{0},
& 
&\chi_{+\a}^\dag(-\vp)\chi_{-\b}(-\vp-\vq)\Ket{0}.
}
Here, we have taken advantage of the translational invariance of the theory to distinguish states by their total external momentum $\vq$, where $\vq=2\pi\(n_x,n_y\)/L$, $n_{x,y}\in\Z$. Provided the internal momentum $\vp$ is not such that $\vp+\vq=-\vp$, these states are distinct for each $\a,\,\b$, and have energy
\eq{\label{eqn:2particleE}
E_f(\vq,\vp)&=\abs{\vp+\vq}+\abs{\vp}.
}
Na\"{i}vely counting, for every $\vq$ and $\vp$, the flavour symmetry gives (at least) $2N_f^2$ such states (additional degeneracies may be present depending on the lattice and internal momentum, but this will not be important for the subsequent discussion).  
When $\vp+\vq=-\vp$, the two states in Eq.~\ref{eqn:2particle} are identical, and there are only $N_f^2$ possible states.

This story no longer holds even at $N_f=\infty$.
The gauge field only couples to single trace operators, so it is natural to expect that the corresponding states may be shifted like in the O($N$) model \cite{Whitsitt2016}. 
However, QED$_3$ and the Dirac-CS theory differ from this example by having four different single-trace fermion bilinear operators: the ``mass" operator $M(x)=\bpsi_\a\psi_\a(x)$ and the global gauge currents, $J^\m(x)=\bpsi_\a\g^\m\psi_\a(x)$. 
It is apparent that the current operators and the mass operator must be treated very differently when we consider the equations of motion:
\eq{\label{eqn:CurrentId}
J^\m&={k\o4\pi}J^\m_{\text{top}}+{i\o e^2}\ep^{\m\n\r}\ptl_\n J_{\text{top},\r},
}
where $J_\text{top}^\m=\ep^{\m\n\r}\ptl_\n A_\r$ is the current of the topological U(1)$_\text{top}$ symmetry. 
This symmetry is equivalent to the non-compactness of $A_\m$ and the irrelevance of monopoles at the fixed point. 
At $N_f=\infty$, when $k=0$, $J_\m$ is more correctly understood as a descendant of the topological current and not as a composite operator. 
In the $e^2\rightarrow\infty$ limit, it vanishes altogether and should not be included in the spectrum: 
all states corresponding the poles of $\Braket{J^\m(x)J^\n(0)}$ in the free theory no longer exist in large-$N_f$ QED$_3$. 
The degeneracy is reduced so that for each total momentum $\vq$ and internal momentum $\vp$ (where $\vp+\vq\neq-\vq$), QED$_3$ has only $2N^2_f-1$ free-fermion-like states with energy $E_f(\vq,\vp)$ (when $\vp+\vq=-\vp$, the degeneracy is further reduced to $N_f^2-1$). This is discussed in more detail in Appendix~\ref{app:Spec}. 
For a small set of momenta, these energy levels are shown in Table~\ref{tab:k0FreeFermEnergies} along with their respective degeneracies.

{\def\arraystretch{0.8}%  1
\begin{table}
\centering
{%\rowcolors{3}{gray!25}{}
\begin{tabular}{ | c | c || c | c  || c | c |}
%\multicolumn{12}{c}{External momentum $\bar{\vq}$}\\\hline
\hline
\multicolumn{2}{|c||}{$\bar{\vq}=(0,0)$} & \multicolumn{2}{c||}{$\bar{\vq}=(1,0)$} & \multicolumn{2}{c|}{$\bar{\vq}=(1,1)$} \\ \hline
$\bar{E}_f$ & $d_f$  & $\bar{E}_f$ & $d_f$& $\bar{E}_f$ &$d_f$   \\\hline\hline
1.414214 & $4N_f^2-2$ & 1.414214 & $2N_f^2-1$ & 1.414214 &  $N_f^2-1$  \\
  & & 2.288246 & $4N_f^2-2$ & 2.288246 & $4N_f^2-2$  \\
  & & & & 2.828427 & $2N_f^2-1$\\
 3.162278 & $8N_f^2-4$ &  3.162278 & $2N_f^2-1$ & 3.162278 & $2N_f^2-1$   \\
  & &   3.702459 & $4N_f^2-2$ & &  \\
  & & 4.130649 & $4N_f^2-2$ & 4.130649 & $4N_f^2-2$   \\
 4.242640 & $4N_f^2-2$ & & & &\\
     & & & & 4.496615 & $4N_f^2-2$ \\
 & & & & 4.670830 & $4N_f^2-2$  \\\hline
  %\hline
% & &  & &  5.036796 & 4 & & & &  & &  \\
% 5.099020 & 8 & & & 5.099020 & 2 &  & && & & \\
%  & &  & &  5.464985 & 4 & & & & & &  \\
% & & 5.489309 & 2 & & & & & & & & \\
\end{tabular}}
\caption{Energies of two-particle fermion states in QED$_3$ (CS level $k=0$) on a square torus of size $L$. 
Energies are shown for $\vq=0$, $\vq_1=2\pi(1,0)/L$ and $\vq_2=2\pi(1,1)/L$. 
The 1st, 3rd, and 5th columns list the energy levels, $E_f$, while the column to the right, labelled $d_f$, shows the degeneracy of the level. 
The energy levels with finite external momentum, $\vq_1=2\pi(1,0)/L$ and $\vq_2=2\pi(1,1)/L$, have an additional 4-fold degeneracy resulting from the symmetry of the lattice. 
($\bar{\vq}=L\vq/2\pi$, $\bar{E}=LE/2\pi$.)}\label{tab:k0FreeFermEnergies}
\end{table}}

For non-vanishing $k$, the situation is very similar. Eq.~\ref{eqn:CurrentId} indicates that the CS term attaches $k$ units of charge to each unit of magnetic flux so that the charged state with the lowest energy has $k$ fermions accompanied by a single unit of magnetic flux. 
In the limit $k\rightarrow\infty$, these states have very high energies and, as in the $k=0$ case, will not contribute to the low energy spectrum. 
The same free-fermion states whose energies are given in Table~\ref{tab:k0FreeFermEnergies} also appear in the Dirac-CS theory with the same degeneracy theory regardless of the level $k$.
%It follows that the same set of energy levels presented in Table~\ref{tab:k0Energies} 
%As in the $k=0$, 

For both QED$_3$ and Dirac-CS, the removal of $J^\m$ is counterbalanced by the addition of $A_\m$. 
The spectrum must be supplemented by the poles of the photon propagator, $\Delta_{\m\n}(x)=\Braket{A_\m(x)A_\n(0)}$, and, unlike for the free-fermion states, the energies of the photon states depend on the level $k$.

From the effective action in Eq.~\ref{eqn:Seff}, the photon propagator is obtained by inverting the polarization matrix $\Pi^{\m\n}(q)$.  
However, as discussed in the previous section, gauge invariance is only fully taken into account once the FP determinant's contribution is included as well. 
Analogous to our identification of $\g_\vq$ as the frequencies in a set of harmonic oscillators for the Maxwell-Chern-Simons theory in Eq.~\ref{eqn:FmcsFreq}, the physical photon modes are actually given by the zeros of the argument of the logarithms in $F_\Gg$. 
When $N_f=\infty$, each mode represents an infinite tower of states of a harmonic oscillator like in Maxwell-Chern-Simons: additional energy levels are present as integer multiples of the modes determined from $F_\Gg$. 
Eqs.~\ref{eqn:Fq0} and \ref{eqn:Fqn0} indicate that these modes occur when the functions
\eq{\label{eqn:Kpoles}
K_0(\w)&=-\(\Pi_f^{xx}(\w,0)\o\w\)^2+{1\o 4\pi^2\lam^2}
\\
 K_\vq(\w)&={\Pi^{00}(\w,\vq)\o\vq^2}\(\Pi^{xx}(\w,\vq)+\Pi^{yy}(\w,\vq)+{\w^2\o\vq^2}\Pi^{00}(\w,\vq)\)-{1\o \vq^4}\(q_y\Pi^{0x}(\w,\vq)-q_x\Pi^{0y}(\w,\vq)\)^2\notag
}
vanish. Here, we have analytically continued to real frequencies, $\w=i\ep$. In what follows $\ep$ will always denote an imaginary frequency, while $\w$ will represent a real frequency; the same symbol for the polarization $\Pi^{\m\n}$ is used for both.
For $k=0$, some modes levels are listed in Table~\ref{tab:k0Energies} while Table~\ref{tab:k0EnergyLevels} shows the lowest energy levels when multi-photon states are included. Table~\ref{tab:kfiniteEnergies} gives the lowest ten modes with zero external momentum for several values of $\lam=N_f/k$.
%Note that integer multiples of the listed energies are also allowed, because each zero of the photon propagator represents an infinite tower  of states of a harmonic oscillator.
{\def\arraystretch{0.65}%  1
\begin{table}
\centering
\begin{tabular}{| c | c || c | c || c | c |}
%\multicolumn{12}{c}{External momentum $\bar{\vq}$}\\\hline
\hline
\multicolumn{2}{|c||}{$\bar{\vq}=(0,0)$} & \multicolumn{2}{c||}{$\bar{\vq}=(1,0)$} & \multicolumn{2}{c|}{$\bar{\vq}=(1,1)$} \\ \hline
 $\bar{E}_\g$ & $d^E_\g$ &$\bar{E}_\g$ &$d^E_\g$ & $\bar{E}_\g$ & $d^E_\g$ \\\hline\hline
 0.58413	&	2	&		&		&		&		\\
1.16826	&	4	&		&		&		&		\\
	&		&	1.43798	&	1	&		&		\\
	&		&		&		&	1.68208	&	1	\\
	&		&		&		&	1.73907	&	1	\\
1.75239	&	8	&		&		&		&		\\
	&		&	1.97629	&	1	&		&		\\
	&		&	2.02211	&	2	&		&		\\
	&		&		&		&	2.26621	&	2	\\
2.31153	&	2	&		&		&		&		\\
	&		&		&		&	2.3232	&	2	\\
2.33652	&	16	&		&		&		&		\\
	&		&		&		&	2.52761	&	1	\\
	&		&	2.56042	&	2	&		&		\\
	&		&	2.60624	&	4	&		&		\\
	&		&	2.65809	&	1	&		&		\\
	&		&		&		&	2.81322	&	1	\\
	&		&		&		&	2.85034	&	4	\\
2.87596	&	2	&		&		&		&		\\
2.89566	&	4	&		&		&		&		\\
2.89566	&	4	&		&		&		&		\\
	&		&		&		&	2.90733	&	4	\\
2.92065	&	32	&		&		&		&		\\
	&		&		&		&	3.11174	&	2	\\
	&		&	3.14455	&	4	&		&		\\
	&		&	3.15634	&	1	&		&		\\
	&		&	3.19037	&	8	&		&		\\
	&		&	3.24222	&	2	&		&		\\
3.36416	&	2	&		&		&		&		\\
	&		&		&		&	3.39735	&	2	\\\hline
\end{tabular}
\caption{Photon energy levels in QED$_3$ (CS level $k=0$) on a square torus of size $L$. 
Energies are shown for states with total momentum $\vq=0$, $\vq_1=2\pi(1,0)/L$ and $\vq_2=2\pi(1,1)/L$. 
The 1st, 3rd, and 5th columns list the energy levels, $E_\g$, while the column  immediately to the right provides their degeneracy, $d^{E}_\g$. 
($\bar{\vq}=L\vq/2\pi$, $\bar{E}=LE/2\pi$.)}\label{tab:k0EnergyLevels}
\end{table}}

To summarize, the $N_f=\infty$ theory does not have single-particle excitations. Instead, the lowest energy states are of the form given in Eq.~\ref{eqn:2particle} or are created by the photon, $A_\m$. 
The free fermion 2-particle energies $E_f(\vq,\vp)$ occur with either a (2$N_f^2-1$) or a {$(N_f^2-1)$-fold} degeneracy depending on the internal momentum $\vp$ (and before additional lattice symmetries are taken into account). 
The frequency modes of the photon operator are given by the gauge-fixed poles of $\Delta_{\m\n}$ and correspond to the zeros of the expressions in Eq.~\ref{eqn:Kpoles}. Each mode, $\w_\g$, represents a harmonic oscillator so that the energies $2\w_\g, 3\w_\g,\w_\g+\w_{\g'}, \dots$ are present in the spectrum as well.
We will examine Eq.~\ref{eqn:Kpoles} in more detail in the subsequent sections. 

\subsection{Zero external momentum, $\vq=0$}
\begin{figure}
\centering
\includegraphics[scale=0.75]{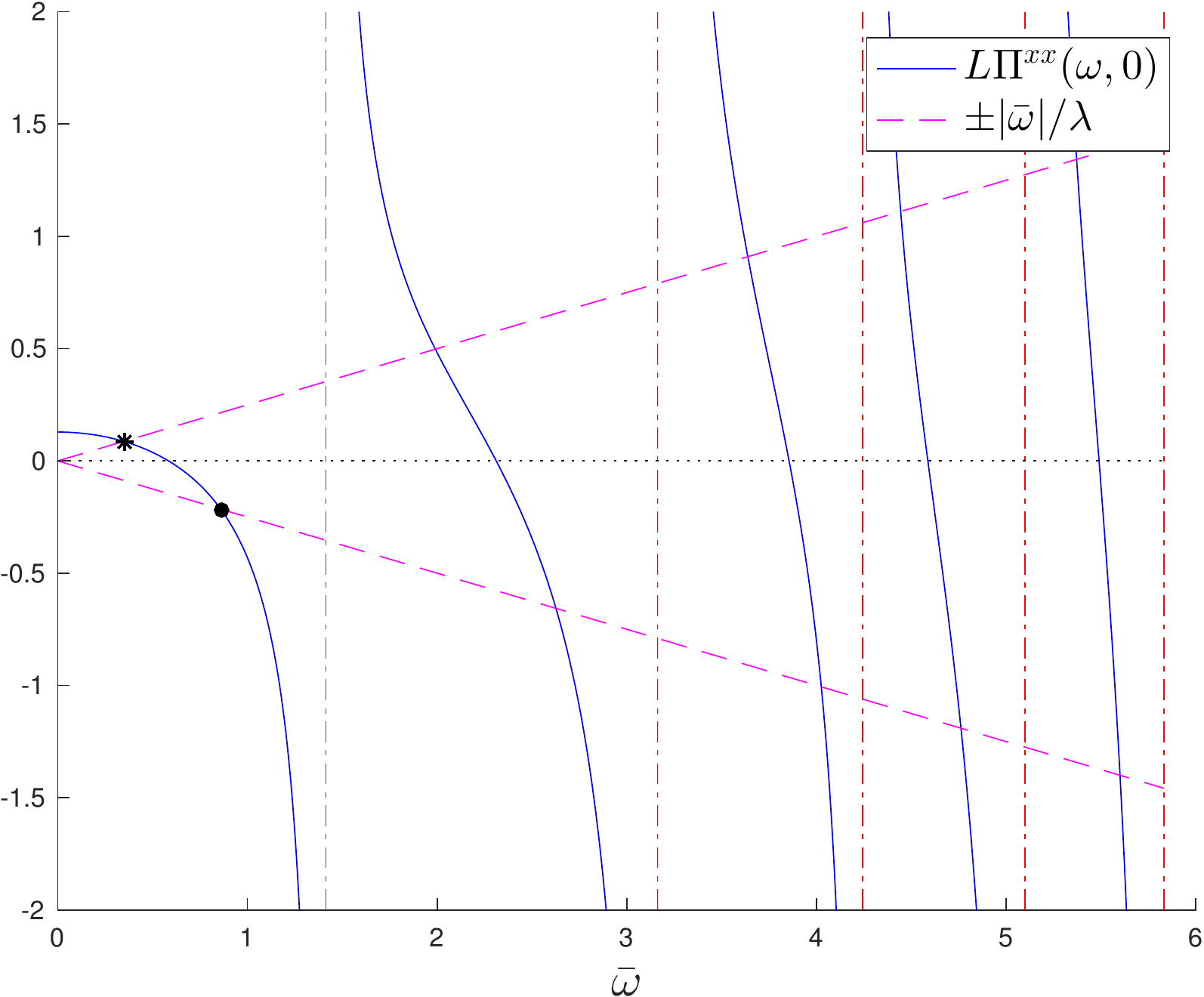}
\caption{Plot of $\Pi_f^{xx}(\w,0)$ and $\abs{\w}/2\pi\lam$. When $k=0$, the modes are two-fold degenerate and occur when $\Pi_f^{xx}=0$. For $k\neq0$, the degeneracy splits and the frequencies are given by the intersection points $\Pi^{xx}_f(\w,0)=\pm \abs{\w}/(2\pi\lam)$. For $\lam=4$, this occurs when the solid blue and dashed magenta lines cross. The lowest and second-lowest energies are shown in black with an asterisk and a circle respectively. The vertical dash-dotted lines in red mark the poles of $\Pi^{xx}_f$ at the two-particle energies of the free theory. ($\bar{\w}=L\w/2\pi$.)}
\label{fig:g0}
\end{figure}

When the external momentum vanishes, the zeros of Eq.~\ref{eqn:Kpoles} occur when
\eq{\label{eqn:q0En}
\Pi^{xx}_f(\w,0)&=\pm{\abs{\w}\o2\pi\lam}.
}
In Fig.~\ref{fig:g0}, the left-hand side is shown with a solid blue line and the right-hand side is shown with a dashed magenta line for $\lam=4$.

When $k=0$ ($\lam\rightarrow\infty$), the energy modes are two-fold degenerate and are given by the point where $\Pi^{xx}_f$ crosses the $x$-axis. This degeneracy may be surprising since in 2+1 dimensions we expect the photon to have a single polarization. 
However, if we had approached the problem by gauge fixing in the Coulomb gauge, we would immediately see that the constraint $\mathbf{\nabla}\cdot\v{A}=0$ does not affect the $\vq=0$ modes, again resulting in a degeneracy. 
In fact, the exact degeneracy is a result of the additional symmetry of our torus, which gives $\Pi^{xx}_f(\ep,0)=\Pi^{yy}_f(\ep,0)$.

To understand the effect of the gauge field on the theory, it's useful to explicitly write the form $\Pi_f^{xx}(\w,0)$ takes:
\eq{\label{eqn:PiXXfq0}
\Pi^{xx}_f(\w,0)&={y_2\o4\pi L}- {\w^2\o 2 L^2}\sum_\vp {1\o\abs{\vp}}{1\o 4\vp^2-\w^2}
}
where  $y_2=-Y_2(1/2)\cong1.6156$ for the function $Y_2(s)$ defined in Eq.~\ref{eqn:Y2}. 
Schematically, we see from Fig.~\ref{fig:g0} that we could rewrite this as a rational function:
\eq{
\Pi^{xx}_f(\w,0)\sim {\prod_\g \(\w^2-\w_\g^2\)\o\prod_\vp\(\w^2-E_f(0,\vp)^2\)}
}
where $\w_\g$ are the zeros of the polarization, $\Pi^{xx}_f(\w_\g,0)=0$, and $E_f(0,\vp)=2\abs{\vp}$ are its poles. 
Its contribution to the partition function is therefore something like
\eq{
Z^{\vq=0}\sim \prod_{i\ep_n}\left\{{\prod_\vp\[\(i\ep_n\)^2+4\vp^2\]\o\prod_\g \[\(i\ep_n\)^2+\(\w_\g\)^2\]}\right\}^2.
}
Not only are the interacting theory's energies present as poles, but the free theory's two-particle energies are accounted for as zeros in the numerator, thereby removing them from the spectral function. The fact that the function is squared accounts for the square symmetry of the torus. % degeneracy of $J^x$ and $J^y$ at $\vq=0$ in the free theory.

\begin{figure}
\centering
\includegraphics[scale=0.75]{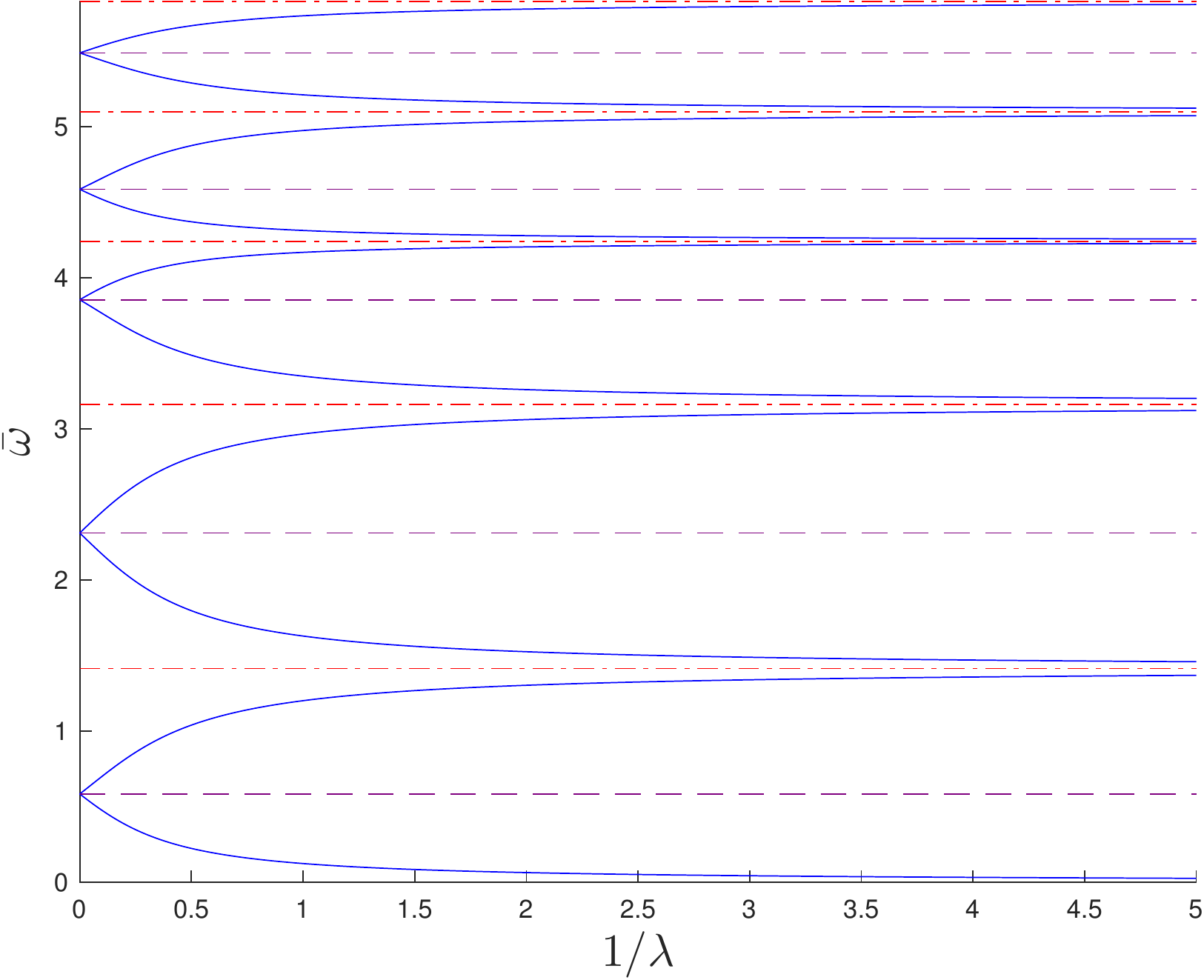}
\caption{Plot of the modes of the Dirac-CS theory as a function of $1/\lambda$. When $1/\lam\rightarrow0$, the  CS term vanishes, and the energies are two-fold degenerate, occurring when $\Pi_f^{xx}=0$. These are marked with the dashed purple line. As $1/\lambda$ becomes large, the lowest mode approaches zero and all others approach the two-particle energies of the free theory, shown with a dash-dotted red line.  ($\bar{\w}=L\w/2\pi$, $\lam=N_f/k$.)}
\label{fig:lambdaVSenergy}
\end{figure}

When $k$ is non-zero, the degeneracy splits. The energies are depicted in Fig.~\ref{fig:g0} as the intersection points of $\Pi^{xx}_f$ and $\pm\abs{\w}/2\pi\lam$ for $\lam=4$. 
Fig.~\ref{fig:lambdaVSenergy} plots the first few modes in blue as a function of $1/\lam$, and for several values of $\lam$, the first ten modes are listed in Table~\ref{tab:kfiniteEnergies}. 
When $\lam$ is very large, these modes have only a small splitting and are nearly the same as in QED$_3$, shown with the purple dashed line in Fig.~\ref{fig:lambdaVSenergy}. 
Conversely, as $\lam\rightarrow0$, the lowest mode $\w_0^*$ approaches zero while all other levels approach one of the free theory two-particle energies, depicted with a dash-dotted red line in Fig.~\ref{fig:lambdaVSenergy}. 

The lowest energy level, $\w^*_0$, can be identified as the splitting between the groundstates of the pure CS theory induced by matter. 
In the limit of $\lam$ and $\w^*_0$ very small, the topological degeneracy is restored  (albeit in the $k\rightarrow\infty$ limit).
This aligns with out expectation that gauge fluctuations are suppressed at large $k$ even when $N_f$ is small \cite{Chen1993}.
In a similar fashion, when the fermions have a large mass $M_f$, we find $\lim_{\w\rightarrow0}\Pi^{xx}_f(\w,0)\sim e^{-M_f}$, once again implying an effective topological ground-state degeneracy.

\subsection{Finite external momentum, $\vq\neq0$}

\begin{table}
%\centering
{\rowcolors{3}{gray!25}{}
\begin{tabular}{| c | c | c | c | c| c | c | c|}
\multicolumn{8}{c}{$\lam=N_f/k$}\\\hline
0 & 1/10 & 1/4 & 1/2 & 1 & 4 & 10 & $\infty$ \\\hline\hline
%\multicolumn{2}{|c||}{$1/4$} & \multicolumn{2}{c||}{$1/2$} & \multicolumn{2}{c||}{$1$}  & \multicolumn{2}{c||}{$4$}& \multicolumn{2}{c|}{$10$}\\\hline
0 & 0.012851 & 0.032056 &   0.063615& 0.123519 &  0.347859 &  0.475391 & 0.584130\\
 & 1.39173 & 1.358213&1.303479&1.201486& 0.859690 &0.700684& \\
1.4142136 & 1.436722 & 1.470375 &  1.525588&   1.629405&  1.990723 &  2.171077 & 2.311525\\ 
 & 3.142113 & 3.111848  &3.061891&2.966946&2.626458 &2.450844&\\
 3.162278& 3.182355 & 3.212169 &  3.260552& 3.349688&  3.637930 &  3.765391& 3.855225\\
 & 4.235129 & 4.223855 &4.205187 & 4.169170& 4.025093 &3.935641 &\\
4.242641 & 4.250129 & 4.261281    &  4.279522& 4.313961 &  4.443737 & 4.761364 & 4.586816 \\
 & 5.086480 & 5.067543 &5.036016&4.975471 & 4.519975&4.660037&\\
5.099020 & 5.111437 & 5.129740   &   5.159072& 5.211794 &   5.371116 & 5.439288 & 5.489309 \\
 & 5.820132 & 5.804317 &5.779259&5.734850& 5.599761 &5.537818&\\\hline
\end{tabular}}
\caption{Dirac-Chern-Simons modes at $N_f,\,k=\infty$ with zero external momentum, $\vq=0$. ($\bar{\w}=LE/2\pi$.)}\label{tab:kfiniteEnergies}
\end{table}

The situation for finite external momenta is very similar. Using Eq.~\ref{eqn:Kpoles}, along with Eqs.~\ref{eqn:Pixx} and \ref{eqn:Pixy}, all levels can be numerically evaluated for any value of $\lam$.

\begin{figure}
\centering
\includegraphics[scale=0.75]{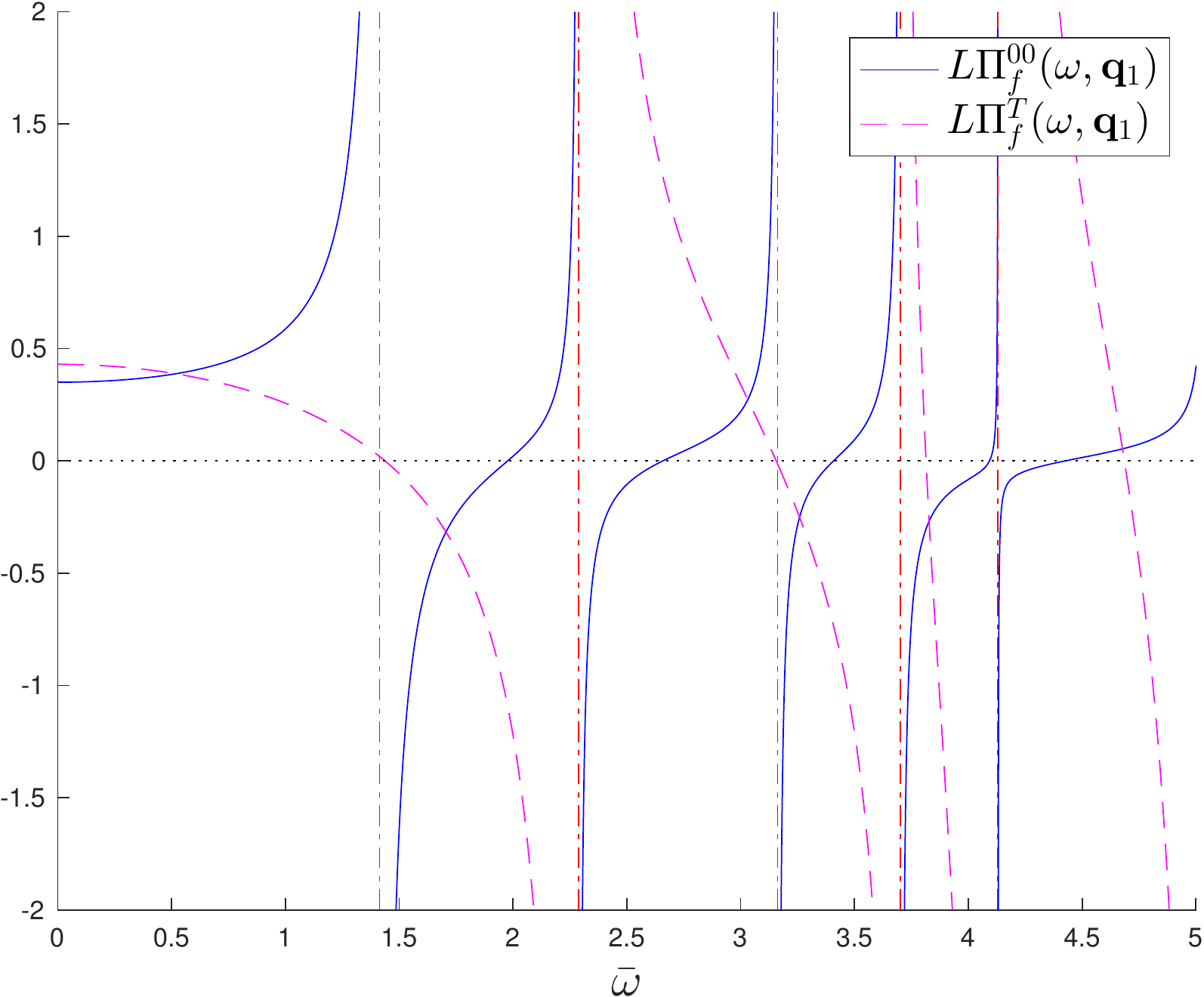}
\caption{Plot of $\Pi_f^{00}(\w,\vq_1)$ and $\Pi_f^T(\w,\vq_1)$ for $\vq_1=2\pi(1,0)/L$, shown in solid blue and dashed magenta respectively. The vertical dash-dotted lines in red denote the two-particle energies of the free theory, $E_f(\vq_1,\vp)$. ($\bar{\w}=L\w/2\pi$.)}
\label{fig:Pi0TQ10}
\end{figure}

The next-lowest energies occur when the total momentum is $\vq_1=2\pi\(1,0\)/L$, or any other of the momenta related to it by a $\pi/2$ rotation: $2\pi(0,1)/L$, $2\pi(-1,0)/L$, and $2\pi(0,-1)/L$.
The C$_4$ symmetry of the square torus implies an additional four-fold degeneracy for all energy levels which would not generally be present. 
For these particular momenta, it turns out that the second term of $K_{\vq_1}(\w)$ in Eq.~\ref{eqn:Kpoles} vanishes for all $\w$ when $k=0$, and the zeros of the determinant can be found by separately solving for the zeros of $\Pi^{00}_f$ and $\Pi^T_f=\Pi^{xx}_f+\Pi^{yy}_f+\w^2\Pi^{00}_f/\vq^2$. These functions are plotted in Fig.~\ref{fig:Pi0TQ10} and the resulting modes are given in Table~\ref{tab:k0Energies} along with the results for $\vq_2=2\pi(1,1)/L$.

\section{Conclusion}\label{sec:conclusion}

This paper has described the structure of 2+1 dimensional conformal gauge theories on the two-torus $\v{T}^2$. We computed the partition
function on $\v{T}^2 \times \mathds{R}$ in the limit of large fermion flavor number, $N_f$, 
using strategies similar to those employed for the computation on the three-sphere $\v{S}^3$ in Ref.~\onlinecite{Klebanov2012}.
We also deduced the energies of the low-lying states in the spectrum. For large $N_f$, most of the states are simply given by
the sum of the free fermion energies with anti-periodic boundary conditions, as established in Appendix~\ref{app:freeFermion}. 
However, singlet combinations of pairs of fermions which couple to the current operator
are strongly renormalized even at $N_f = \infty$: these states appear instead as bound states given by the zeros of the effective
action for the gauge field. A similar phenomenon appears \cite{Whitsitt2016} in the O($N$) Wilson-Fisher conformal theory.

These results should be useful in identifying possible realizations of non-trivial conformal field theories in exact diagonalization
studies of model quantum spin systems.

\subsection*{Acknowledgements}

We thank S. Pufu and S. Whitsitt for useful discussions.
The research was supported by the NSF under Grant DMR-1360789.
Research at Perimeter Institute is supported by the Government of Canada through Industry Canada and by the Province of Ontario 
through the Ministry of Research and Innovation. SS also acknowledges support from Cenovus Energy at Perimeter Institute. AT is supported by NSERC.

\appendix

\section{Generalized Epstein zeta function}\label{app:Y2}
We define the function $Y_2(s)$ to be
\eq{
Y_2(s)&=\sum_{n_1,n_2=-\infty}^\infty\[\(n_1+{1\o2}\)^2+\(n_2+{1\o2}\)^2\]^{-s}.
\label{eqn:Y2}
}
It is only convergent for $\Re s>1/2$, but can be defined by analytically continuing outside of this domain. 
Specifically, it can be expressed in terms of the special functions $\lam$ and $\b$ \cite{Zucker1974}:
\eq{
Y_2(s)=4\cdot2^s\,\lam(s)\b(s),
}
where
\eq{
\b(s)&=\sum_{n=0}^\infty\(-1\)^n\(2n+1\)^{-s},
&
\lam(s)&=\sum_{n=0}^\infty(2n+1)^{-s}=(1-2^{-s})\zeta(s)
}
with $\zeta(s)=\sum_{n=1}^\infty n^{-s}$, the Riemann zeta function.

\section{Analytic continuation of Maxwell-Chern-Simons free energy}\label{app:MaxCS}

In Eq.~\ref{eqn:EpsteinZeta} we expressed the summation over imaginary frequencies in terms of the Epstein zeta function
\eq{
\zeta_{\mathcal{E}}(s;a^2)&=\sum_{n=-\infty}^\infty \[ n^2+a^2\]^{-s},
}
where $a=\b\g_\vq/2\pi$. This expression is only valid for $\Re s>1/2$, but can be analytically continued onto the entire complex plane. To see this, we use the identity
\eq{
{1\o A^s}&={\pi^s\o\Gamma(s)}\int_0^\infty dt\, t^{s-1}e^{-\pi t A},
}
to write
\eq{
\zeta_{\mathcal{E}}(s;a^2)&=\sum_n{\pi^s\o\Gamma(s)}\int_0^\infty dt\, t^{s-1}e^{-\pi t\( n^2+a^2\)}.
}
For sufficiently large values of $s$, we can exchange the summation and the integral, and, subsequently, use the Poisson summation formula:
\eq{
\zeta_{\mathcal{E}}(s;a^2)&={\pi^s\o\Gamma(s)}\int_0^\infty dt\, t^{s-1}e^{-\pi t a^2}\sum_ne^{-\pi tn^2}
={\pi^s\o\Gamma(s)}\int_0^\infty dt\, t^{s-1}e^{-\pi t a^2}{1\o\sqrt{t}}\sum_\ell e^{-\pi \ell^2/t}.
}
We see that divergence for $\Re s\leq1/2$ is due to the $\ell=0$ term in the sum. Separating this term out and evaluating the integral, we have
\eq{
\zeta_{\mathcal{E}}(s;a^2)&=a^{1-2s}{\sqrt{\pi}\Gamma(s-1/2)\o\Gamma(s)}+{2\pi^s\o\Gamma(s)}\sum_{\ell=1}^\infty \int_0^\infty dt\,t^{s-3/2}e^{-\pi a^2 t}e^{-\pi \ell^2/t}.
}
We can now extend $s$ all the way to zero. Taking the derivative and limit, we have
\eq{
-\lim_{s\rightarrow0}{d\o ds}\zeta_{\mathcal{E}}(s;a^2)&=2\pi a-2\sum_{\ell=1}^\infty {e^{-2\pi a \ell}\o\ell}
=2\pi a + 2\log\(1-e^{-2\pi a}\).
}
Plugging this result into Eq.~\ref{eqn:FmcsSum}, we obtain
\eq{
F_{\mathrm{MCS}}&=-{1\o\b}\log k -{1\o\b}\sum_\vq\log\[ e^{-\b\g_\vq/2}\o1-e^{-\b\g_\vq}\].
}

\section{Leading order contribution}\label{app:freeFermion}
The leading order contribution in the zero temperature limit is
\eq{
F_0(\bb{a})&=-{1\o\b}\sum_p\log\(p+a\)^2=-\sum_\vp\int{d\w\o2\pi}\log\(\w^2+\(\vp+\bb{a}\)^2\)
\nt
&=
-\sum_\vp\int {d\w\o2\pi}\log\w^2-\sum_\vp\abs{\vp+\bb{a}},
}
where $p=(\w,\vp)$, $\vp=2\pi(n_x+1/2,n_y+1/2)/L$, $\(n_x,n_y\)\in\Z^2$.
The first term vanishes using zeta-reg and the second one can be evaluating by analytically continuing to arbitrary $s$:
\eq{
F_0(\bb{a})&=-\sum_\vp\(\vp+\v{a}\)^{-2s}=-N_f\(2\pi \o L\)^{-2s}\sum_\vn\(\vn+{1\o2}+\bb{\a}\)^{-2s}
}
where
\eq{
\a_\m&={L\o2\pi}a_\m.
}
We can write this as
\eq{
F_0(\bb{a})&=-\(2\pi\o L\)^{-2s}{\pi^s\o \Gamma(s)}\[ {1\o s-1}+\int_1^\infty dt \,t^{s-1}\Theta\begin{bmatrix}\bb{\a}\\ 0\end{bmatrix}\(it\)
+  \int_1^\infty dt \,t^{-s}\(\Theta\begin{bmatrix}0\\\bb{\a}\end{bmatrix}\(it\)-1\)\]
}
where $\Theta$ is shorthand for a product of Jacobi theta functions
\eq{
\Theta\begin{bmatrix}\bb{\a}\\ 0\end{bmatrix}\(it\)&=\prod_{j=1,2}\jthetac{\a_j+1/2}{0}(0|it),
&
\Theta\begin{bmatrix}0\\\bb{\a}\end{bmatrix}\(it\)&=\prod_{j=1,2}\jthetac{0}{-\a_j-1/2}(0|it).
}
and we've used the following definition for the Jacobi theta functions with characteristics:
\eq{
\vartheta \begin{bmatrix} a \\ b \end{bmatrix}(\n|\t)
&=
\exp\[ \pi i a^2\t+2\pi i a(\n+b)\]\vartheta(\n+a\t+b|\t)
\nt
&=\sum_{n=-\infty}^\infty \exp\[\pi i (n+a)^2\t+2\pi i (n+a)(\n+b)\].
}
For $s=-1/2$, we have
\eq{
F_0(\bb{a})&={1\o L}\[ -{2\o3}+ \int_1^\infty dt \,t^{-3/2} \Theta\begin{bmatrix}\bb{\a}\\ 0\end{bmatrix}\(it\)
+\int_1^\infty dt\,\sqrt{t}\(\Theta\begin{bmatrix}0\\-\bb{\a}\end{bmatrix}\(it\)-1\)\].
}
This function is plotted in Fig.~\ref{fig:F0vsa} and clearly has a minimum at $\bb{\a}=\(0,0\)$. In terms of the function $Y_2$ defined in Appendix~\ref{app:Y2} in Eq.~\ref{eqn:Y2}, this
\eq{
F_0(0)&=-{2\pi\o L}Y_2\(-{1\o2}\).
}
\begin{figure}
\centering
\includegraphics[scale=0.6]{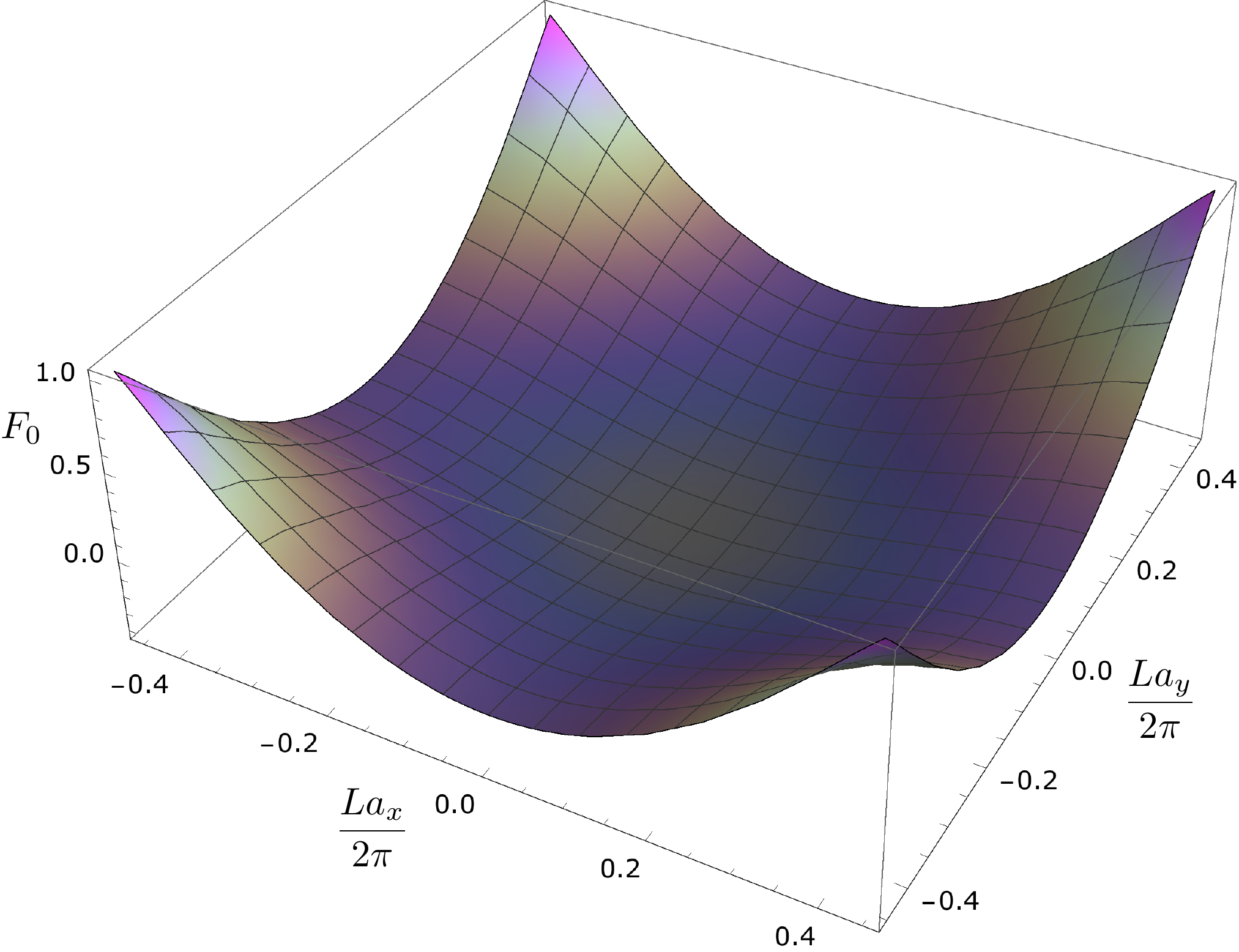}
\caption{Plot on the free energy of a free Dirac fermion on the torus as a function of its boundary conditions, $a_x$, $a_y$.}
\label{fig:F0vsa}
\end{figure}

\section{Polarization diagram}\label{app:Pif}

Here we calculate the leading $1/N_f$ contribution to the gauge kernel from the fermions. It is given by the polarization diagram:
\eq{
S_f[B]&={1\o2}\tr\( {1\o i\sd{\ptl}}\,\sd{B}\,{1\o i\sd{\ptl}}\,\sd{B}\)
\nt
&={1\o2}{1\o\b V}\sum_{p,q}\tr\( {\sd{p}\o p^2}\sd{B}(-q)\,{\(\sd{p}+\sd{q}\)\o\(p+q\)^2}\,\sd{B}(q)\)
\nt
&={1\o2}{1\o\b V}\sum_{p,q}\tr\(\s^\r\s^\m\s^\lam\s^\n\)B_\m(-q)B_\n(q){p_\r\(p+q\)_\lam\o p^2\(p+q\)^2}
}
where we have dropped all explicit references to $\bar{a}=0$. The internal momentum, $p$, corresponds to a fermionic field, $p^\m=2\pi(n^\m+1/2)/L_\m$, $n_\m\in\Z$, whereas the external momentum is appropriate for a bosonic field, $q^\m=2\pi n^\m/L_\m$, $n^\m\in\Z$. This can be written as
\eq{\label{eqn:Polar}
S_f[B]&={1\o2}\sum_q B_\n(-q)\Pi_f^{\m\n}(q)B_\n(q)
}
with
\eq{
\Pi_f^{\m\n}(q)&={1\o\b V}\sum_{p}\tr\(\s^\r\s^\m\s^\lam\s^\n\){p_\r\(p+q\)_\lam\o p^2\(p+q\)^2}
\nt
%&={2\o L^2}\sum_\vp\int{d\w\o 2\pi} \[ { 2p^\m p^\n - \d^{\m\n}p^2\o  p^2\(p+q\)^2}+{ p^\m q^\n + q^\m p^\n - \d^{\m\n}p\cdot q\o  p^2\(p+q\)^2}\].
%\nt
&={2\o\b L^2}\sum_p{ p^\m\(p^\n+q^\n\)+\(p^\m+q^\m\)p^\n-\d^{\m\n}p\cdot \(p+q\) \o p^2\(p+q\)^2}.
}
In what follows, we will consider the zero temperature limit, $\b\rightarrow\infty$.

We begin by calculating the the $xx$ component:
\eq{
\Pi_f^{xx}(\ep,\vq)&={2\o L^2}\sum_\vp\int{d\w\o2\pi}{ p_x(p_x+q_x)-p_y(p_y+q_y)-\w(\w+\ep)\o\(\w^2+\vp^2\)\(\(\w+\ep\)^2+\(\vp+\vq\)^2\)}
\nt
&={1\o L^2}\sum_\vp\[{p_x(p_x+q_x)-p_y(p_y+q_y)\o \(\abs{\vp}+\abs{\vp+\vq}\)^2+\ep^2}\({1\o\abs{\vp}}+{1\o\abs{\vp+\vq}}\)-{\abs{\vp}+\abs{\vp+\vq}\o\(\abs{\vp}+\abs{\vp+\vq}\)^2+\ep^2}\]
}
This is formally divergent but can be regulated by adding and subtracting the divergent piece and analytically continuing using zeta functions:
\eq{
\Pi_f^{xx}(\ep,\vq)&=-{1\o L^2}\Bigg\{\sum_\vp{p_x(p_x+q_x)-p_y(p_y+q_y)\o \(\abs{\vp}+\abs{\vp+\vq}\)^2+\ep^2}\({1\o\abs{\vp}}+{1\o\abs{\vp+\vq}}\)
\nt
&\quad-\sum_\vp\[{\abs{\vp}+\abs{\vp+\vq}\o\(\abs{\vp}+\abs{\vp+\vq}\)^2+\ep^2}-{1\o 2\abs{\vp}}\]+\sum_\vp{1\o 2\abs{\vp}}\Bigg\}.
}
The divergent term is
\eq{
\sum_\vp{1\o 2\abs{\vp}}&={1\o2}{L\o 2\pi}\sum_\vn {1\o \sqrt{ \(\vn+1/2\)^2}}={L\o 4\pi}Y_2\(1/2\),
}
where $Y_2(s)$ is defined for all $s$ in Appendix~\ref{app:Y2}.
The finite expression is therefore
\eq{\label{eqn:Pixx}
\Pi_f^{xx}(\ep,\vq)&=-{1\o 4\pi L}Y_2(1/2)-{1\o L^2}\sum_\vp\[{\abs{\vp}+\abs{\vp+\vq}\o\(\abs{\vp}+\abs{\vp+\vq}\)^2+\ep^2}-{1\o 2\abs{\vp}}\]
\nt
&\quad+{1\o L^2}\Bigg\{\sum_\vp{p_x(p_x+q_x)-p_y(p_y+q_y)\o \(\abs{\vp}+\abs{\vp+\vq}\)^2+\ep^2}\({1\o\abs{\vp}}+{1\o\abs{\vp+\vq}}\).
}
Similarly, we find
\eq{
\Pi_f^{yy}(\ep,\vq)&=-{1\o 4\pi L}Y_2(1/2)-{1\o L^2}\sum_\vp\[{\abs{\vp}+\abs{\vp+\vq}\o\(\abs{\vp}+\abs{\vp+\vq}\)^2+\ep^2}-{1\o 2\abs{\vp}}\]
\nt
&\quad-{1\o L^2}\Bigg\{\sum_\vp{p_x(p_x+q_x)-p_y(p_y+q_y)\o \(\abs{\vp}+\abs{\vp+\vq}\)^2+\ep^2}\({1\o\abs{\vp}}+{1\o\abs{\vp+\vq}}\),
\nt
\Pi_f^{xy}(\ep,\vq)&={1\o L^2} \sum_\vp{p_x(p_y+q_y)+p_y(p_x+q_x)\o \(\abs{\vp}+\abs{\vp+\vq}\)^2+\ep^2}
\({1\o\abs{\vp}}+{1\o\abs{\vp+\vq}}\),
\nt
\Pi_f^{00}(\ep,\vq)&={1\o L^2}\sum_\vp{\abs{\vp}\abs{\vp+\vq}-\vp\cdot\(\vp+\vq\)\o \(\abs{\vp}+\abs{\vp+\vq}\)^2+\ep^2}\({1\o\abs{\vp}}+{1\o\abs{\vp+\vq}}\),
\nt
\Pi_f^{0i}(\ep,\vq)&=  {1\o L^2}\sum_\vp{\ep\o \(\abs{\vp}+\abs{\vp+\vq}\)^2+\ep^2}\({p_i\o\abs{\vp}}-{p_i+q_i\o\abs{\vp+\vq}}\).
\label{eqn:Pixy}}

\section{Operator contributions to the spectrum}\label{app:Spec}

In Sec.~\ref{sec:Spectrum} we stated that in addition to imposing charge-neutrality, the gauge field alters the spectrum in two ways at $N_f=\infty$. First, its presence enforces the constraint $J^\m(x)=0$, removing one state from the spectrum for every choice of external momentum $\vq$ and internal momentum $\vp$, thereby decreasing the degeneracy of the free theory spectrum. 
Further, the photon creates states which contribute to the spectrum as well; their energies coincide with the
poles of the photon propagator, $\Delta_{\m\n}(x)=\Braket{A_\m(x)A_\n(0)}$. 

We can understand how this comes about by translating the field theoretic operators to the quantum mechanical language of the free theory. We write
\eq{
J^\mu(x)&={1\o L^2}\sum_{\vq,E} e^{-i\vx\cdot\vq} J_E^\m(\vq),
& 
J_E^\m(\vq)&=\sum_{\substack{\vp\\E_f(\vq,\vp)=E}} \bpsi_\a(\vp+\vq)\g^\m\psi_\a(\vp),
\nt
M(x)&={1\o L^2}\sum_{\vq,E} e^{-i\vx\cdot\vq} M_E(\vq),
&
M_E(\vq)&=\sum_{\substack{\vp\\E_f(\vq,\vp)=E}} \bpsi_\a(\vp+\vq)\psi_\a(\vp).
}
For the moment, we specify to the case where $\vp+\vq\neq-\vp$. Eq.~\ref{eqn:2particle} shows the two distinct states which exist for each energy $E_f(\vq,\vp)$ (additional degeneracies may be present due to the symmetry of the lattice, but this does not alter any of the following discussion). 
It follows that $J_E^\m(\vq)$ and $M_E(\vq)$ create states of the form
\eq{
J^\m_E(\vq)\Ket{0}&=\[v_1^\m(\vp)\chi_{+\a}^\dag(\vp+\vq)\chi_{-\a}(\vp)+v_2^\m(\vp)\chi^\dag_{+\a}(-\vp)\chi_{-\a}(-\vp-\vq)\]\Ket{0},
\nt
M_E(\vq)\Ket{0}&=\[v_1^M(\vp)\chi_{+\a}^\dag(\vp+\vq)\chi_{-\a}(\vp)+v_2^M(\vp)\chi^\dag_{+\a}(-\vp)\chi_{-\a}(-\vp-\vq)\]\Ket{0},
}
where the ``$E$'' subscript on $v^\m_i(\vp)$ and $v^M_i(\vp)$ has been dropped for notational ease. 
These coefficients are easily computed, and are found to be
\eq{\label{eqn:JMcoeff}
\bb{v}^0&={i\o2}\begin{pmatrix}1-{P\o\abs{\vp}}{\overline{P+Q}\o\abs{\vp+\vq}}\\ 1-{\overline{P}\o\abs{\vp}}{{P+Q}\o\abs{\vp+\vq}}\end{pmatrix},
&
\bb{v}^x&={1\o2}\begin{pmatrix}-{P\o\abs{\vp}}+{\overline{P+Q}\o\abs{\vp+\vq}}\\-{\overline{P}\o\abs{\vp}}+{{P+Q}\o\abs{\vp+\vq}}\end{pmatrix},
&
\bb{v}^y&={i\o2}\begin{pmatrix}{P\o\abs{\vp}}+{\overline{P+Q}\o\abs{\vp+\vq}}\\-{\overline{P}\o\abs{\vp}}-{{P+Q}\o\abs{\vp+\vq}}\end{pmatrix},
\nt
\bb{v}^M&={i\o2}\begin{pmatrix}1+{P\o\abs{\vp}}{\overline{P+Q}\o\abs{\vp+\vq}}\\1+{\overline{P}\o\abs{\vp}}{{P+Q}\o\abs{\vp+\vq}}\end{pmatrix},
}
where $P=p_x+ip_y$, $Q=q_x+iq_y$. 
While it may not be obvious, it can be verified that the state created by the mass operator is orthogonal to the three states created by the current operators, and that these states are all proportional to one another. 

The linear dependence of the current states actually follows directly from the conservation law $\ptl_\m J^\m=0$. In terms of the states, this reads
\eq{\label{eqn:JCons}
\[-i\(\abs{\vp+\vq}+\abs{\vp}\)J^0_E(\vq)+q_x J_E^x(\vq)+q_y J^y_E(\vq)\]\Ket{0}=0.
}
The space spanned by $\chi_{+\a}^\dag(\vp+\vq)\chi_{-\a}(\vp)\Ket{0}$ and $\chi^\dag_{+\a}(-\vp)\chi_{-\a}(-\vp-\vq)\Ket{0}$ is a 2-dimensional complex vector space, equivalent to a 4$d$ \emph{real} vector space. 
Eq.~\ref{eqn:JCons} shows that the three $J^\m_E(\vq)\Ket{0}$ states actually only span a 2$d$ real subspace, \emph{ie.} a 1$d$ \emph{complex} vector space. As claimed, the currents only create a single state. 
The orthogonality of $M_E(\vq)\Ket{0}$ to this state is then obvious since Eq.~(\ref{eqn:JMcoeff}) implies that
\eq{(\bb{v}^\mu)^\dag \bb{v}^M=0.} 
Returning to the large-$N_f$ theory, the gauge current states cease to exist, but the mass state remains, resulting in (at least) a $2N_f^2-1$ degeneracy.
  
In the special case $\vp+\vq=-\vp$, there is only a single state for each $\a,\b$ pair, and so only a $N_f^2$ degeneracy in the free theory.
Eq.~\ref{eqn:JMcoeff} shows that only the current operators create states of this form, and, as above, this state is removed at $N_f=\infty$, resulting in a $N_f^2-1$ degeneracy.

\bibliographystyle{apsrev4-1_custom}
\bibliography{torusRef}

\begin{thebibliography}{44}%
\makeatletter
\providecommand \@ifxundefined [1]{%
 \@ifx{#1\undefined}
}%
\providecommand \@ifnum [1]{%
 \ifnum #1\expandafter \@firstoftwo
 \else \expandafter \@secondoftwo
 \fi
}%
\providecommand \@ifx [1]{%
 \ifx #1\expandafter \@firstoftwo
 \else \expandafter \@secondoftwo
 \fi
}%
\providecommand \natexlab [1]{#1}%
\providecommand \enquote  [1]{``#1''}%
\providecommand \bibnamefont  [1]{#1}%
\providecommand \bibfnamefont [1]{#1}%
\providecommand \citenamefont [1]{#1}%
\providecommand \href@noop [0]{\@secondoftwo}%
\providecommand \href [0]{\begingroup \@sanitize@url \@href}%
\providecommand \@href[1]{\@@startlink{#1}\@@href}%
\providecommand \@@href[1]{\endgroup#1\@@endlink}%
\providecommand \@sanitize@url [0]{\catcode `\\12\catcode `\$12\catcode
  `\&12\catcode `\#12\catcode `\^12\catcode `\_12\catcode `\%12\relax}%
\providecommand \@@startlink[1]{}%
\providecommand \@@endlink[0]{}%
\providecommand \url  [0]{\begingroup\@sanitize@url \@url }%
\providecommand \@url [1]{\endgroup\@href {#1}{\urlprefix }}%
\providecommand \urlprefix  [0]{URL }%
\providecommand \Eprint [0]{\href }%
\providecommand \doibase [0]{http://dx.doi.org/}%
\providecommand \selectlanguage [0]{\@gobble}%
\providecommand \bibinfo  [0]{\@secondoftwo}%
\providecommand \bibfield  [0]{\@secondoftwo}%
\providecommand \translation [1]{[#1]}%
\providecommand \BibitemOpen [0]{}%
\providecommand \bibitemStop [0]{}%
\providecommand \bibitemNoStop [0]{.\EOS\space}%
\providecommand \EOS [0]{\spacefactor3000\relax}%
\providecommand \BibitemShut  [1]{\csname bibitem#1\endcsname}%
\let\auto@bib@innerbib\@empty
%</preamble>
\bibitem [{\citenamefont {{Schuler}}\ \emph {et~al.}(2016)\citenamefont
  {{Schuler}}, \citenamefont {{Whitsitt}}, \citenamefont {{Henry}},
  \citenamefont {{Sachdev}},\ and\ \citenamefont
  {{L{\"a}uchli}}}]{Schuler2016}%
  \BibitemOpen
  \bibfield  {author} {\bibinfo {author} {\bibfnamefont {M.}~\bibnamefont
  {{Schuler}}}, \bibinfo {author} {\bibfnamefont {S.}~\bibnamefont
  {{Whitsitt}}}, \bibinfo {author} {\bibfnamefont {L.-P.}\ \bibnamefont
  {{Henry}}}, \bibinfo {author} {\bibfnamefont {S.}~\bibnamefont {{Sachdev}}},
  \ and\ \bibinfo {author} {\bibfnamefont {A.~M.}\ \bibnamefont
  {{L{\"a}uchli}}},\ }\bibfield  {title} {\enquote {\bibinfo {title}
  {{Universal Signatures of Quantum Critical Points from Finite-Size Torus
  Spectra: A Window into the Operator Content of Higher-Dimensional Conformal
  Field Theories}},}\ }\href@noop {} {\bibfield  {journal} {\bibinfo  {journal}
  {ArXiv e-prints}\ } (\bibinfo {year} {2016})},\ \Eprint
  {http://arxiv.org/abs/1603.03042} {arXiv:1603.03042 [cond-mat.str-el]}
  \BibitemShut {NoStop}%
\bibitem [{\citenamefont {{Whitsitt}}\ and\ \citenamefont
  {{Sachdev}}(2016)}]{Whitsitt2016}%
  \BibitemOpen
  \bibfield  {author} {\bibinfo {author} {\bibfnamefont {S.}~\bibnamefont
  {{Whitsitt}}}\ and\ \bibinfo {author} {\bibfnamefont {S.}~\bibnamefont
  {{Sachdev}}},\ }\bibfield  {title} {\enquote {\bibinfo {title} {{Transition
  from the $\mathbb{Z}_{2}$ spin liquid to antiferromagnetic order: Spectrum on
  the torus}},}\ }\href {\doibase 10.1103/PhysRevB.94.085134} {\bibfield
  {journal} {\bibinfo  {journal} {Phys. Rev. B}\ }\textbf {\bibinfo {volume}
  {94}},\ \bibinfo {eid} {085134} (\bibinfo {year} {2016})},\ \Eprint
  {http://arxiv.org/abs/1603.05652} {arXiv:1603.05652 [cond-mat.str-el]}
  \BibitemShut {NoStop}%
\bibitem [{\citenamefont {{Rantner}}\ and\ \citenamefont
  {{Wen}}(2001)}]{Rantner2001}%
  \BibitemOpen
  \bibfield  {author} {\bibinfo {author} {\bibfnamefont {W.}~\bibnamefont
  {{Rantner}}}\ and\ \bibinfo {author} {\bibfnamefont {X.-G.}\ \bibnamefont
  {{Wen}}},\ }\bibfield  {title} {\enquote {\bibinfo {title} {{Electron
  Spectral Function and Algebraic Spin Liquid for the Normal State of
  Underdoped High T$_{c}$ Superconductors}},}\ }\href {\doibase
  10.1103/PhysRevLett.86.3871} {\bibfield  {journal} {\bibinfo  {journal}
  {Phys. Rev. Lett.}\ }\textbf {\bibinfo {volume} {86}},\ \bibinfo {pages}
  {3871} (\bibinfo {year} {2001})},\ \Eprint
  {http://arxiv.org/abs/cond-mat/0010378} {cond-mat/0010378} \BibitemShut
  {NoStop}%
\bibitem [{\citenamefont {{Wen}}(2002)}]{Wen2002}%
  \BibitemOpen
  \bibfield  {author} {\bibinfo {author} {\bibfnamefont {X.-G.}\ \bibnamefont
  {{Wen}}},\ }\bibfield  {title} {\enquote {\bibinfo {title} {{Quantum orders
  and symmetric spin liquids}},}\ }\href {\doibase 10.1103/PhysRevB.65.165113}
  {\bibfield  {journal} {\bibinfo  {journal} {Phys. Rev. B}\ }\textbf {\bibinfo
  {volume} {65}},\ \bibinfo {eid} {165113} (\bibinfo {year} {2002})},\ \Eprint
  {http://arxiv.org/abs/cond-mat/0107071} {cond-mat/0107071} \BibitemShut
  {NoStop}%
\bibitem [{\citenamefont {{Hermele}}\ \emph {et~al.}(2004)\citenamefont
  {{Hermele}}, \citenamefont {{Senthil}}, \citenamefont {{Fisher}},
  \citenamefont {{Lee}}, \citenamefont {{Nagaosa}},\ and\ \citenamefont
  {{Wen}}}]{Hermele2004}%
  \BibitemOpen
  \bibfield  {author} {\bibinfo {author} {\bibfnamefont {M.}~\bibnamefont
  {{Hermele}}}, \bibinfo {author} {\bibfnamefont {T.}~\bibnamefont
  {{Senthil}}}, \bibinfo {author} {\bibfnamefont {M.~P.~A.}\ \bibnamefont
  {{Fisher}}}, \bibinfo {author} {\bibfnamefont {P.~A.}\ \bibnamefont {{Lee}}},
  \bibinfo {author} {\bibfnamefont {N.}~\bibnamefont {{Nagaosa}}}, \ and\
  \bibinfo {author} {\bibfnamefont {X.-G.}\ \bibnamefont {{Wen}}},\ }\bibfield
  {title} {\enquote {\bibinfo {title} {{Stability of U (1) spin liquids in two
  dimensions}},}\ }\href {\doibase 10.1103/PhysRevB.70.214437} {\bibfield
  {journal} {\bibinfo  {journal} {Phys. Rev. B}\ }\textbf {\bibinfo {volume}
  {70}},\ \bibinfo {eid} {214437} (\bibinfo {year} {2004})},\ \Eprint
  {http://arxiv.org/abs/cond-mat/0404751} {cond-mat/0404751} \BibitemShut
  {NoStop}%
\bibitem [{\citenamefont {Karthik}\ and\ \citenamefont
  {Narayanan}(2016{\natexlab{a}})}]{Karthik15}%
  \BibitemOpen
  \bibfield  {author} {\bibinfo {author} {\bibfnamefont {N.}~\bibnamefont
  {Karthik}}\ and\ \bibinfo {author} {\bibfnamefont {R.}~\bibnamefont
  {Narayanan}},\ }\bibfield  {title} {\enquote {\bibinfo {title} {{No evidence
  for bilinear condensate in parity-invariant three-dimensional QED with
  massless fermions}},}\ }\href {\doibase 10.1103/PhysRevD.93.045020}
  {\bibfield  {journal} {\bibinfo  {journal} {Phys. Rev. D}\ }\textbf {\bibinfo
  {volume} {93}},\ \bibinfo {pages} {045020} (\bibinfo {year}
  {2016}{\natexlab{a}})},\ \Eprint {http://arxiv.org/abs/1512.02993}
  {arXiv:1512.02993 [hep-lat]} \BibitemShut {NoStop}%
%%CITATION = ARXIV:1512.02993;%%
\bibitem [{\citenamefont {Karthik}\ and\ \citenamefont
  {Narayanan}(2016{\natexlab{b}})}]{Karthik16}%
  \BibitemOpen
  \bibfield  {author} {\bibinfo {author} {\bibfnamefont {N.}~\bibnamefont
  {Karthik}}\ and\ \bibinfo {author} {\bibfnamefont {R.}~\bibnamefont
  {Narayanan}},\ }\bibfield  {title} {\enquote {\bibinfo {title}
  {{Scale-invariance of parity-invariant three-dimensional QED}},}\ }\href@noop
  {} {\  (\bibinfo {year} {2016}{\natexlab{b}})},\ \Eprint
  {http://arxiv.org/abs/1606.04109} {arXiv:1606.04109 [hep-th]} \BibitemShut
  {NoStop}%
%%CITATION = ARXIV:1606.04109;%%
\bibitem [{\citenamefont {{Yan}}\ \emph {et~al.}(2011)\citenamefont {{Yan}},
  \citenamefont {{Huse}},\ and\ \citenamefont {{White}}}]{Yan11}%
  \BibitemOpen
  \bibfield  {author} {\bibinfo {author} {\bibfnamefont {S.}~\bibnamefont
  {{Yan}}}, \bibinfo {author} {\bibfnamefont {D.~A.}\ \bibnamefont {{Huse}}}, \
  and\ \bibinfo {author} {\bibfnamefont {S.~R.}\ \bibnamefont {{White}}},\
  }\bibfield  {title} {\enquote {\bibinfo {title} {{Spin-Liquid Ground State of
  the S = 1/2 Kagome Heisenberg Antiferromagnet}},}\ }\href {\doibase
  10.1126/science.1201080} {\bibfield  {journal} {\bibinfo  {journal}
  {Science}\ }\textbf {\bibinfo {volume} {332}},\ \bibinfo {pages} {1173}
  (\bibinfo {year} {2011})},\ \Eprint {http://arxiv.org/abs/1011.6114}
  {arXiv:1011.6114 [cond-mat.str-el]} \BibitemShut {NoStop}%
\bibitem [{\citenamefont {{Jiang}}\ \emph {et~al.}(2012)\citenamefont
  {{Jiang}}, \citenamefont {{Wang}},\ and\ \citenamefont
  {{Balents}}}]{Jiang12}%
  \BibitemOpen
  \bibfield  {author} {\bibinfo {author} {\bibfnamefont {H.-C.}\ \bibnamefont
  {{Jiang}}}, \bibinfo {author} {\bibfnamefont {Z.}~\bibnamefont {{Wang}}}, \
  and\ \bibinfo {author} {\bibfnamefont {L.}~\bibnamefont {{Balents}}},\
  }\bibfield  {title} {\enquote {\bibinfo {title} {{Identifying topological
  order by entanglement entropy}},}\ }\href {\doibase 10.1038/nphys2465}
  {\bibfield  {journal} {\bibinfo  {journal} {Nature Physics}\ }\textbf
  {\bibinfo {volume} {8}},\ \bibinfo {pages} {902} (\bibinfo {year} {2012})},\
  \Eprint {http://arxiv.org/abs/1205.4289} {arXiv:1205.4289 [cond-mat.str-el]}
  \BibitemShut {NoStop}%
\bibitem [{\citenamefont {{Depenbrock}}\ \emph {et~al.}(2012)\citenamefont
  {{Depenbrock}}, \citenamefont {{McCulloch}},\ and\ \citenamefont
  {{Schollwoeck}}}]{McCulloch12}%
  \BibitemOpen
  \bibfield  {author} {\bibinfo {author} {\bibfnamefont {S.}~\bibnamefont
  {{Depenbrock}}}, \bibinfo {author} {\bibfnamefont {I.~P.}\ \bibnamefont
  {{McCulloch}}}, \ and\ \bibinfo {author} {\bibfnamefont {U.}~\bibnamefont
  {{Schollwoeck}}},\ }\bibfield  {title} {\enquote {\bibinfo {title} {{Nature
  of the Spin Liquid Ground State of the S=1/2 Kagome Heisenberg Model}},}\
  }\href@noop {} {\bibfield  {journal} {\bibinfo  {journal} {ArXiv e-prints}\ }
  (\bibinfo {year} {2012})},\ \Eprint {http://arxiv.org/abs/1205.4858}
  {arXiv:1205.4858 [cond-mat.str-el]} \BibitemShut {NoStop}%
\bibitem [{\citenamefont {{Zhu}}\ and\ \citenamefont
  {{White}}(2015)}]{ZhuWhite15}%
  \BibitemOpen
  \bibfield  {author} {\bibinfo {author} {\bibfnamefont {Z.}~\bibnamefont
  {{Zhu}}}\ and\ \bibinfo {author} {\bibfnamefont {S.~R.}\ \bibnamefont
  {{White}}},\ }\bibfield  {title} {\enquote {\bibinfo {title} {{Spin liquid
  phase of the S =1/2 J$_{1}$-J$_{2}$ Heisenberg model on the triangular
  lattice}},}\ }\href {\doibase 10.1103/PhysRevB.92.041105} {\bibfield
  {journal} {\bibinfo  {journal} {Phys. Rev. B}\ }\textbf {\bibinfo {volume}
  {92}},\ \bibinfo {eid} {041105} (\bibinfo {year} {2015})},\ \Eprint
  {http://arxiv.org/abs/1502.04831} {arXiv:1502.04831 [cond-mat.str-el]}
  \BibitemShut {NoStop}%
\bibitem [{\citenamefont {{Saadatmand}}\ and\ \citenamefont
  {{McCulloch}}(2016)}]{McCulloch16}%
  \BibitemOpen
  \bibfield  {author} {\bibinfo {author} {\bibfnamefont {S.~N.}\ \bibnamefont
  {{Saadatmand}}}\ and\ \bibinfo {author} {\bibfnamefont {I.~P.}\ \bibnamefont
  {{McCulloch}}},\ }\bibfield  {title} {\enquote {\bibinfo {title} {{Symmetry
  fractionalization in the topological phase of the spin-1/2 J1-J2 triangular
  Heisenberg model}},}\ }\href@noop {} {\bibfield  {journal} {\bibinfo
  {journal} {ArXiv e-prints}\ } (\bibinfo {year} {2016})},\ \Eprint
  {http://arxiv.org/abs/1606.00334} {arXiv:1606.00334 [cond-mat.str-el]}
  \BibitemShut {NoStop}%
\bibitem [{\citenamefont {{Mei}}\ \emph {et~al.}(2016)\citenamefont {{Mei}},
  \citenamefont {{Chen}}, \citenamefont {{He}},\ and\ \citenamefont
  {{Wen}}}]{Mei16}%
  \BibitemOpen
  \bibfield  {author} {\bibinfo {author} {\bibfnamefont {J.-W.}\ \bibnamefont
  {{Mei}}}, \bibinfo {author} {\bibfnamefont {J.-Y.}\ \bibnamefont {{Chen}}},
  \bibinfo {author} {\bibfnamefont {H.}~\bibnamefont {{He}}}, \ and\ \bibinfo
  {author} {\bibfnamefont {X.-G.}\ \bibnamefont {{Wen}}},\ }\bibfield  {title}
  {\enquote {\bibinfo {title} {{SU(2) spin-rotation symmetric tensor network
  state for spin-1/2 Heisenberg model on kagome lattice and its modular
  matrices}},}\ }\href@noop {} {\bibfield  {journal} {\bibinfo  {journal}
  {ArXiv e-prints}\ } (\bibinfo {year} {2016})},\ \Eprint
  {http://arxiv.org/abs/1606.09639} {arXiv:1606.09639 [cond-mat.str-el]}
  \BibitemShut {NoStop}%
\bibitem [{\citenamefont {{Iqbal}}\ \emph {et~al.}(2015)\citenamefont
  {{Iqbal}}, \citenamefont {{Poilblanc}},\ and\ \citenamefont
  {{Becca}}}]{Iqbal14}%
  \BibitemOpen
  \bibfield  {author} {\bibinfo {author} {\bibfnamefont {Y.}~\bibnamefont
  {{Iqbal}}}, \bibinfo {author} {\bibfnamefont {D.}~\bibnamefont
  {{Poilblanc}}}, \ and\ \bibinfo {author} {\bibfnamefont {F.}~\bibnamefont
  {{Becca}}},\ }\bibfield  {title} {\enquote {\bibinfo {title} {{Spin-1/2
  Heisenberg J$_{1}$-J$_{2}$ antiferromagnet on the kagome lattice}},}\ }\href
  {\doibase 10.1103/PhysRevB.91.020402} {\bibfield  {journal} {\bibinfo
  {journal} {Phys. Rev. B}\ }\textbf {\bibinfo {volume} {91}},\ \bibinfo {eid}
  {020402} (\bibinfo {year} {2015})},\ \Eprint {http://arxiv.org/abs/1410.7359}
  {arXiv:1410.7359 [cond-mat.str-el]} \BibitemShut {NoStop}%
\bibitem [{\citenamefont {{Iqbal}}\ \emph {et~al.}(2016)\citenamefont
  {{Iqbal}}, \citenamefont {{Hu}}, \citenamefont {{Thomale}}, \citenamefont
  {{Poilblanc}},\ and\ \citenamefont {{Becca}}}]{Iqbal16}%
  \BibitemOpen
  \bibfield  {author} {\bibinfo {author} {\bibfnamefont {Y.}~\bibnamefont
  {{Iqbal}}}, \bibinfo {author} {\bibfnamefont {W.-J.}\ \bibnamefont {{Hu}}},
  \bibinfo {author} {\bibfnamefont {R.}~\bibnamefont {{Thomale}}}, \bibinfo
  {author} {\bibfnamefont {D.}~\bibnamefont {{Poilblanc}}}, \ and\ \bibinfo
  {author} {\bibfnamefont {F.}~\bibnamefont {{Becca}}},\ }\bibfield  {title}
  {\enquote {\bibinfo {title} {{Spin liquid nature in the Heisenberg
  J$_{1}$-J$_{2}$ triangular antiferromagnet}},}\ }\href {\doibase
  10.1103/PhysRevB.93.144411} {\bibfield  {journal} {\bibinfo  {journal} {Phys.
  Rev. B}\ }\textbf {\bibinfo {volume} {93}},\ \bibinfo {eid} {144411}
  (\bibinfo {year} {2016})},\ \Eprint {http://arxiv.org/abs/1601.06018}
  {arXiv:1601.06018 [cond-mat.str-el]} \BibitemShut {NoStop}%
\bibitem [{\citenamefont {{Grover}}\ and\ \citenamefont
  {{Vishwanath}}(2013)}]{Grover2013}%
  \BibitemOpen
  \bibfield  {author} {\bibinfo {author} {\bibfnamefont {T.}~\bibnamefont
  {{Grover}}}\ and\ \bibinfo {author} {\bibfnamefont {A.}~\bibnamefont
  {{Vishwanath}}},\ }\bibfield  {title} {\enquote {\bibinfo {title} {{Quantum
  phase transition between integer quantum Hall states of bosons}},}\ }\href
  {\doibase 10.1103/PhysRevB.87.045129} {\bibfield  {journal} {\bibinfo
  {journal} {Phys. Rev. B}\ }\textbf {\bibinfo {volume} {87}},\ \bibinfo {eid}
  {045129} (\bibinfo {year} {2013})},\ \Eprint {http://arxiv.org/abs/1210.0907}
  {arXiv:1210.0907 [cond-mat.str-el]} \BibitemShut {NoStop}%
\bibitem [{\citenamefont {Barkeshli}(2013)}]{Barkeshli2013}%
  \BibitemOpen
  \bibfield  {author} {\bibinfo {author} {\bibfnamefont {M.}~\bibnamefont
  {Barkeshli}},\ }\bibfield  {title} {\enquote {\bibinfo {title} {{Transitions
  between chiral spin liquids and $\mathbb{Z}_2$ spin liquids}},}\ }\href
  {http://arxiv.org/abs/1307.8194} {\bibfield  {journal} {\bibinfo  {journal}
  {arXiv preprint arXiv:1307.8194}\ } (\bibinfo {year} {2013})}\BibitemShut
  {NoStop}%
\bibitem [{\citenamefont {{Senthil}}\ \emph
  {et~al.}(2004{\natexlab{a}})\citenamefont {{Senthil}}, \citenamefont
  {{Vishwanath}}, \citenamefont {{Balents}}, \citenamefont {{Sachdev}},\ and\
  \citenamefont {{Fisher}}}]{Senthil2004}%
  \BibitemOpen
  \bibfield  {author} {\bibinfo {author} {\bibfnamefont {T.}~\bibnamefont
  {{Senthil}}}, \bibinfo {author} {\bibfnamefont {A.}~\bibnamefont
  {{Vishwanath}}}, \bibinfo {author} {\bibfnamefont {L.}~\bibnamefont
  {{Balents}}}, \bibinfo {author} {\bibfnamefont {S.}~\bibnamefont
  {{Sachdev}}}, \ and\ \bibinfo {author} {\bibfnamefont {M.~P.~A.}\
  \bibnamefont {{Fisher}}},\ }\bibfield  {title} {\enquote {\bibinfo {title}
  {{Deconfined Quantum Critical Points}},}\ }\href {\doibase
  10.1126/science.1091806} {\bibfield  {journal} {\bibinfo  {journal}
  {Science}\ }\textbf {\bibinfo {volume} {303}},\ \bibinfo {pages} {1490}
  (\bibinfo {year} {2004}{\natexlab{a}})},\ \Eprint
  {http://arxiv.org/abs/cond-mat/0311326} {cond-mat/0311326} \BibitemShut
  {NoStop}%
\bibitem [{\citenamefont {{Senthil}}\ \emph
  {et~al.}(2004{\natexlab{b}})\citenamefont {{Senthil}}, \citenamefont
  {{Balents}}, \citenamefont {{Sachdev}}, \citenamefont {{Vishwanath}},\ and\
  \citenamefont {{Fisher}}}]{Senthil2004b}%
  \BibitemOpen
  \bibfield  {author} {\bibinfo {author} {\bibfnamefont {T.}~\bibnamefont
  {{Senthil}}}, \bibinfo {author} {\bibfnamefont {L.}~\bibnamefont
  {{Balents}}}, \bibinfo {author} {\bibfnamefont {S.}~\bibnamefont
  {{Sachdev}}}, \bibinfo {author} {\bibfnamefont {A.}~\bibnamefont
  {{Vishwanath}}}, \ and\ \bibinfo {author} {\bibfnamefont {M.~P.~A.}\
  \bibnamefont {{Fisher}}},\ }\bibfield  {title} {\enquote {\bibinfo {title}
  {{Quantum criticality beyond the Landau-Ginzburg-Wilson paradigm}},}\ }\href
  {\doibase 10.1103/PhysRevB.70.144407} {\bibfield  {journal} {\bibinfo
  {journal} {Phys. Rev. B}\ }\textbf {\bibinfo {volume} {70}},\ \bibinfo {eid}
  {144407} (\bibinfo {year} {2004}{\natexlab{b}})},\ \Eprint
  {http://arxiv.org/abs/cond-mat/0312617} {cond-mat/0312617} \BibitemShut
  {NoStop}%
\bibitem [{\citenamefont {{Klebanov}}\ \emph {et~al.}(2012)\citenamefont
  {{Klebanov}}, \citenamefont {{Pufu}}, \citenamefont {{Sachdev}},\ and\
  \citenamefont {{Safdi}}}]{Klebanov2012}%
  \BibitemOpen
  \bibfield  {author} {\bibinfo {author} {\bibfnamefont {I.~R.}\ \bibnamefont
  {{Klebanov}}}, \bibinfo {author} {\bibfnamefont {S.~S.}\ \bibnamefont
  {{Pufu}}}, \bibinfo {author} {\bibfnamefont {S.}~\bibnamefont {{Sachdev}}}, \
  and\ \bibinfo {author} {\bibfnamefont {B.~R.}\ \bibnamefont {{Safdi}}},\
  }\bibfield  {title} {\enquote {\bibinfo {title} {{Entanglement entropy of 3-d
  conformal gauge theories with many flavors}},}\ }\href {\doibase
  10.1007/JHEP05(2012)036} {\bibfield  {journal} {\bibinfo  {journal} {Journal
  of High Energy Physics}\ }\textbf {\bibinfo {volume} {5}},\ \bibinfo {eid}
  {36} (\bibinfo {year} {2012})},\ \Eprint {http://arxiv.org/abs/1112.5342}
  {arXiv:1112.5342 [hep-th]} \BibitemShut {NoStop}%
\bibitem [{\citenamefont {Chen}\ \emph {et~al.}(1993)\citenamefont {Chen},
  \citenamefont {Fisher},\ and\ \citenamefont {Wu}}]{Chen1993}%
  \BibitemOpen
  \bibfield  {author} {\bibinfo {author} {\bibfnamefont {W.}~\bibnamefont
  {Chen}}, \bibinfo {author} {\bibfnamefont {M.~P.~A.}\ \bibnamefont {Fisher}},
  \ and\ \bibinfo {author} {\bibfnamefont {Y.-S.}\ \bibnamefont {Wu}},\
  }\bibfield  {title} {\enquote {\bibinfo {title} {Mott transition in an anyon
  gas},}\ }\href {\doibase 10.1103/PhysRevB.48.13749} {\bibfield  {journal}
  {\bibinfo  {journal} {Phys. Rev. B}\ }\textbf {\bibinfo {volume} {48}},\
  \bibinfo {pages} {13749} (\bibinfo {year} {1993})}\BibitemShut {NoStop}%
\bibitem [{\citenamefont {{Sachdev}}(1998)}]{Sachdev1998}%
  \BibitemOpen
  \bibfield  {author} {\bibinfo {author} {\bibfnamefont {S.}~\bibnamefont
  {{Sachdev}}},\ }\bibfield  {title} {\enquote {\bibinfo {title}
  {{Nonzero-temperature transport near fractional quantum Hall critical
  points}},}\ }\href {\doibase 10.1103/PhysRevB.57.7157} {\bibfield  {journal}
  {\bibinfo  {journal} {Phys. Rev. B}\ }\textbf {\bibinfo {volume} {57}},\
  \bibinfo {pages} {7157} (\bibinfo {year} {1998})},\ \Eprint
  {http://arxiv.org/abs/cond-mat/9709243} {cond-mat/9709243} \BibitemShut
  {NoStop}%
\bibitem [{\citenamefont {{Hickey}}\ \emph {et~al.}(2016)\citenamefont
  {{Hickey}}, \citenamefont {{Cincio}}, \citenamefont {{Papi{\'c}}},\ and\
  \citenamefont {{Paramekanti}}}]{Arun16}%
  \BibitemOpen
  \bibfield  {author} {\bibinfo {author} {\bibfnamefont {C.}~\bibnamefont
  {{Hickey}}}, \bibinfo {author} {\bibfnamefont {L.}~\bibnamefont {{Cincio}}},
  \bibinfo {author} {\bibfnamefont {Z.}~\bibnamefont {{Papi{\'c}}}}, \ and\
  \bibinfo {author} {\bibfnamefont {A.}~\bibnamefont {{Paramekanti}}},\
  }\bibfield  {title} {\enquote {\bibinfo {title} {{Haldane-Hubbard Mott
  Insulator: From Tetrahedral Spin Crystal to Chiral Spin Liquid}},}\ }\href
  {\doibase 10.1103/PhysRevLett.116.137202} {\bibfield  {journal} {\bibinfo
  {journal} {Phys. Rev. Lett.}\ }\textbf {\bibinfo {volume} {116}},\ \bibinfo
  {eid} {137202} (\bibinfo {year} {2016})},\ \Eprint
  {http://arxiv.org/abs/1509.08461} {arXiv:1509.08461 [cond-mat.str-el]}
  \BibitemShut {NoStop}%
\bibitem [{\citenamefont {{Wietek}}\ and\ \citenamefont
  {{L{\"a}uchli}}(2016)}]{Lauchli16}%
  \BibitemOpen
  \bibfield  {author} {\bibinfo {author} {\bibfnamefont {A.}~\bibnamefont
  {{Wietek}}}\ and\ \bibinfo {author} {\bibfnamefont {A.~M.}\ \bibnamefont
  {{L{\"a}uchli}}},\ }\bibfield  {title} {\enquote {\bibinfo {title} {{Chiral
  Spin Liquid and Quantum Criticality in Extended $S=1/2$ Heisenberg Models on
  the Triangular Lattice}},}\ }\href@noop {} {\bibfield  {journal} {\bibinfo
  {journal} {ArXiv e-prints}\ } (\bibinfo {year} {2016})},\ \Eprint
  {http://arxiv.org/abs/1604.07829} {arXiv:1604.07829 [cond-mat.str-el]}
  \BibitemShut {NoStop}%
\bibitem [{\citenamefont {Witten}(1989)}]{Witten1989}%
  \BibitemOpen
  \bibfield  {author} {\bibinfo {author} {\bibfnamefont {E.}~\bibnamefont
  {Witten}},\ }\bibfield  {title} {\enquote {\bibinfo {title} {{Quantum field
  theory and the Jones polynomial}},}\ }\href {\doibase 10.1007/BF01217730}
  {\bibfield  {journal} {\bibinfo  {journal} {Communications in Mathematical
  Physics}\ }\textbf {\bibinfo {volume} {121}},\ \bibinfo {pages} {351}
  (\bibinfo {year} {1989})}\BibitemShut {NoStop}%
\bibitem [{\citenamefont {Wen}\ and\ \citenamefont {Niu}(1990)}]{Wen1990}%
  \BibitemOpen
  \bibfield  {author} {\bibinfo {author} {\bibfnamefont {X.~G.}\ \bibnamefont
  {Wen}}\ and\ \bibinfo {author} {\bibfnamefont {Q.}~\bibnamefont {Niu}},\
  }\bibfield  {title} {\enquote {\bibinfo {title} {{Ground-state degeneracy of
  the fractional quantum Hall states in the presence of a random potential and
  on high-genus Riemann surfaces}},}\ }\href {\doibase
  10.1103/PhysRevB.41.9377} {\bibfield  {journal} {\bibinfo  {journal} {Phys.
  Rev. B}\ }\textbf {\bibinfo {volume} {41}},\ \bibinfo {pages} {9377}
  (\bibinfo {year} {1990})}\BibitemShut {NoStop}%
\bibitem [{\citenamefont {{Hastings}}(2001)}]{Hastings2000}%
  \BibitemOpen
  \bibfield  {author} {\bibinfo {author} {\bibfnamefont {M.~B.}\ \bibnamefont
  {{Hastings}}},\ }\bibfield  {title} {\enquote {\bibinfo {title} {{Dirac
  structure, RVB, and Goldstone modes in the kagom{\'e} antiferromagnet}},}\
  }\href {\doibase 10.1103/PhysRevB.63.014413} {\bibfield  {journal} {\bibinfo
  {journal} {Phys. Rev. B}\ }\textbf {\bibinfo {volume} {63}},\ \bibinfo {eid}
  {014413} (\bibinfo {year} {2001})},\ \Eprint
  {http://arxiv.org/abs/cond-mat/0005391} {cond-mat/0005391} \BibitemShut
  {NoStop}%
\bibitem [{\citenamefont {{Ran}}\ \emph {et~al.}(2007)\citenamefont {{Ran}},
  \citenamefont {{Hermele}}, \citenamefont {{Lee}},\ and\ \citenamefont
  {{Wen}}}]{Ran2007}%
  \BibitemOpen
  \bibfield  {author} {\bibinfo {author} {\bibfnamefont {Y.}~\bibnamefont
  {{Ran}}}, \bibinfo {author} {\bibfnamefont {M.}~\bibnamefont {{Hermele}}},
  \bibinfo {author} {\bibfnamefont {P.~A.}\ \bibnamefont {{Lee}}}, \ and\
  \bibinfo {author} {\bibfnamefont {X.-G.}\ \bibnamefont {{Wen}}},\ }\bibfield
  {title} {\enquote {\bibinfo {title} {{Projected-Wave-Function Study of the
  Spin-1/2 Heisenberg Model on the Kagom{\'e} Lattice}},}\ }\href {\doibase
  10.1103/PhysRevLett.98.117205} {\bibfield  {journal} {\bibinfo  {journal}
  {Phys. Rev. Lett.}\ }\textbf {\bibinfo {volume} {98}},\ \bibinfo {eid}
  {117205} (\bibinfo {year} {2007})},\ \Eprint
  {http://arxiv.org/abs/cond-mat/0611414} {cond-mat/0611414} \BibitemShut
  {NoStop}%
\bibitem [{\citenamefont {{Hermele}}\ \emph {et~al.}(2008)\citenamefont
  {{Hermele}}, \citenamefont {{Ran}}, \citenamefont {{Lee}},\ and\
  \citenamefont {{Wen}}}]{Hermele2008}%
  \BibitemOpen
  \bibfield  {author} {\bibinfo {author} {\bibfnamefont {M.}~\bibnamefont
  {{Hermele}}}, \bibinfo {author} {\bibfnamefont {Y.}~\bibnamefont {{Ran}}},
  \bibinfo {author} {\bibfnamefont {P.~A.}\ \bibnamefont {{Lee}}}, \ and\
  \bibinfo {author} {\bibfnamefont {X.-G.}\ \bibnamefont {{Wen}}},\ }\bibfield
  {title} {\enquote {\bibinfo {title} {{Properties of an algebraic spin liquid
  on the kagome lattice}},}\ }\href {\doibase 10.1103/PhysRevB.77.224413}
  {\bibfield  {journal} {\bibinfo  {journal} {Phys. Rev. B}\ }\textbf {\bibinfo
  {volume} {77}},\ \bibinfo {eid} {224413} (\bibinfo {year} {2008})},\ \Eprint
  {http://arxiv.org/abs/0803.1150} {arXiv:0803.1150 [cond-mat.str-el]}
  \BibitemShut {NoStop}%
\bibitem [{\citenamefont {Polyakov}(1977)}]{Polyakov1977}%
  \BibitemOpen
  \bibfield  {author} {\bibinfo {author} {\bibfnamefont {A.}~\bibnamefont
  {Polyakov}},\ }\bibfield  {title} {\enquote {\bibinfo {title} {Quark
  confinement and topology of gauge theories},}\ }\href {\doibase
  http://dx.doi.org/10.1016/0550-3213(77)90086-4} {\bibfield  {journal}
  {\bibinfo  {journal} {Nuclear Physics B}\ }\textbf {\bibinfo {volume}
  {120}},\ \bibinfo {pages} {429 } (\bibinfo {year} {1977})}\BibitemShut
  {NoStop}%
\bibitem [{\citenamefont {Polyakov}(1987)}]{Polyakov1987}%
  \BibitemOpen
  \bibfield  {author} {\bibinfo {author} {\bibfnamefont {A.~M.}\ \bibnamefont
  {Polyakov}},\ }\href {https://books.google.com/books?id=uaI8xcjJ8LMC} {\emph
  {\bibinfo {title} {Gauge Fields and Strings}}},\ Contemporary concepts in
  physics\ (\bibinfo  {publisher} {Taylor \& Francis},\ \bibinfo {year}
  {1987})\BibitemShut {NoStop}%
\bibitem [{\citenamefont {Borokhov}\ \emph {et~al.}(2002)\citenamefont
  {Borokhov}, \citenamefont {Kapustin},\ and\ \citenamefont
  {Wu}}]{Borokhov2002}%
  \BibitemOpen
  \bibfield  {author} {\bibinfo {author} {\bibfnamefont {V.}~\bibnamefont
  {Borokhov}}, \bibinfo {author} {\bibfnamefont {A.}~\bibnamefont {Kapustin}},
  \ and\ \bibinfo {author} {\bibfnamefont {X.}~\bibnamefont {Wu}},\ }\bibfield
  {title} {\enquote {\bibinfo {title} {{Topological Disorder Operators in
  Three-Dimensional Conformal Field Theory}},}\ }\href
  {http://stacks.iop.org/1126-6708/2002/i=11/a=049} {\bibfield  {journal}
  {\bibinfo  {journal} {Journal of High Energy Physics}\ }\textbf {\bibinfo
  {volume} {2002}},\ \bibinfo {pages} {049} (\bibinfo {year} {2002})},\ \Eprint
  {http://arxiv.org/abs/hep-th/0206054} {arXiv:hep-th/0206054 [hep-th]}
  \BibitemShut {NoStop}%
\bibitem [{\citenamefont {{Pufu}}(2014)}]{Pufu14}%
  \BibitemOpen
  \bibfield  {author} {\bibinfo {author} {\bibfnamefont {S.~S.}\ \bibnamefont
  {{Pufu}}},\ }\bibfield  {title} {\enquote {\bibinfo {title} {{Anomalous
  dimensions of monopole operators in three-dimensional quantum
  electrodynamics}},}\ }\href {\doibase 10.1103/PhysRevD.89.065016} {\bibfield
  {journal} {\bibinfo  {journal} {\prd}\ }\textbf {\bibinfo {volume} {89}},\
  \bibinfo {eid} {065016} (\bibinfo {year} {2014})},\ \Eprint
  {http://arxiv.org/abs/1303.6125} {arXiv:1303.6125 [hep-th]} \BibitemShut
  {NoStop}%
\bibitem [{\citenamefont {{Chester}}\ \emph {et~al.}(2015)\citenamefont
  {{Chester}}, \citenamefont {{Mezei}}, \citenamefont {{Pufu}},\ and\
  \citenamefont {{Yaakov}}}]{Pufu15}%
  \BibitemOpen
  \bibfield  {author} {\bibinfo {author} {\bibfnamefont {S.~M.}\ \bibnamefont
  {{Chester}}}, \bibinfo {author} {\bibfnamefont {M.}~\bibnamefont {{Mezei}}},
  \bibinfo {author} {\bibfnamefont {S.~S.}\ \bibnamefont {{Pufu}}}, \ and\
  \bibinfo {author} {\bibfnamefont {I.}~\bibnamefont {{Yaakov}}},\ }\bibfield
  {title} {\enquote {\bibinfo {title} {{Monopole operators from the
  $4-\epsilon$ expansion}},}\ }\href@noop {} {\bibfield  {journal} {\bibinfo
  {journal} {ArXiv e-prints}\ } (\bibinfo {year} {2015})},\ \Eprint
  {http://arxiv.org/abs/1511.07108} {arXiv:1511.07108 [hep-th]} \BibitemShut
  {NoStop}%
\bibitem [{\citenamefont {Appelquist}\ and\ \citenamefont
  {Pisarski}(1981)}]{Appelquist1981}%
  \BibitemOpen
  \bibfield  {author} {\bibinfo {author} {\bibfnamefont {T.}~\bibnamefont
  {Appelquist}}\ and\ \bibinfo {author} {\bibfnamefont {R.~D.}\ \bibnamefont
  {Pisarski}},\ }\bibfield  {title} {\enquote {\bibinfo {title}
  {{High-temperature Yang-Mills theories and three-dimensional quantum
  chromodynamics}},}\ }\href {\doibase 10.1103/PhysRevD.23.2305} {\bibfield
  {journal} {\bibinfo  {journal} {Phys. Rev. D}\ }\textbf {\bibinfo {volume}
  {23}},\ \bibinfo {pages} {2305} (\bibinfo {year} {1981})}\BibitemShut
  {NoStop}%
\bibitem [{\citenamefont {Jackiw}\ and\ \citenamefont
  {Templeton}(1981)}]{Jackiw1981}%
  \BibitemOpen
  \bibfield  {author} {\bibinfo {author} {\bibfnamefont {R.}~\bibnamefont
  {Jackiw}}\ and\ \bibinfo {author} {\bibfnamefont {S.}~\bibnamefont
  {Templeton}},\ }\bibfield  {title} {\enquote {\bibinfo {title} {{How
  super-renormalizable interactions cure their infrared divergences}},}\ }\href
  {\doibase 10.1103/PhysRevD.23.2291} {\bibfield  {journal} {\bibinfo
  {journal} {Phys. Rev. D}\ }\textbf {\bibinfo {volume} {23}},\ \bibinfo
  {pages} {2291} (\bibinfo {year} {1981})}\BibitemShut {NoStop}%
\bibitem [{\citenamefont {Templeton}(1981{\natexlab{a}})}]{Templeton1981}%
  \BibitemOpen
  \bibfield  {author} {\bibinfo {author} {\bibfnamefont {S.}~\bibnamefont
  {Templeton}},\ }\bibfield  {title} {\enquote {\bibinfo {title} {{Summation of
  dominant coupling constant logarithms in QED$_3$}},}\ }\href {\doibase
  http://dx.doi.org/10.1016/0370-2693(81)90687-0} {\bibfield  {journal}
  {\bibinfo  {journal} {Physics Letters B}\ }\textbf {\bibinfo {volume}
  {103}},\ \bibinfo {pages} {134 } (\bibinfo {year}
  {1981}{\natexlab{a}})}\BibitemShut {NoStop}%
\bibitem [{\citenamefont {Templeton}(1981{\natexlab{b}})}]{Templeton1981b}%
  \BibitemOpen
  \bibfield  {author} {\bibinfo {author} {\bibfnamefont {S.}~\bibnamefont
  {Templeton}},\ }\bibfield  {title} {\enquote {\bibinfo {title} {{Summation of
  coupling-constant logarithms in three-dimensional QED}},}\ }\href {\doibase
  10.1103/PhysRevD.24.3134} {\bibfield  {journal} {\bibinfo  {journal} {Phys.
  Rev. D}\ }\textbf {\bibinfo {volume} {24}},\ \bibinfo {pages} {3134}
  (\bibinfo {year} {1981}{\natexlab{b}})}\BibitemShut {NoStop}%
\bibitem [{\citenamefont {Kalmeyer}\ and\ \citenamefont
  {Laughlin}(1987)}]{Kalmeyer1987}%
  \BibitemOpen
  \bibfield  {author} {\bibinfo {author} {\bibfnamefont {V.}~\bibnamefont
  {Kalmeyer}}\ and\ \bibinfo {author} {\bibfnamefont {R.~B.}\ \bibnamefont
  {Laughlin}},\ }\bibfield  {title} {\enquote {\bibinfo {title} {{Equivalence
  of the resonating-valence-bond and fractional quantum Hall states}},}\ }\href
  {\doibase 10.1103/PhysRevLett.59.2095} {\bibfield  {journal} {\bibinfo
  {journal} {Phys. Rev. Lett.}\ }\textbf {\bibinfo {volume} {59}},\ \bibinfo
  {pages} {2095} (\bibinfo {year} {1987})}\BibitemShut {NoStop}%
\bibitem [{\citenamefont {Polychronakos}(1990)}]{Poly1990}%
  \BibitemOpen
  \bibfield  {author} {\bibinfo {author} {\bibfnamefont {A.~P.}\ \bibnamefont
  {Polychronakos}},\ }\bibfield  {title} {\enquote {\bibinfo {title} {{Abelian
  Chern-Simons theories in 2 + 1 dimensions}},}\ }\href {\doibase
  http://dx.doi.org/10.1016/0003-4916(90)90171-J} {\bibfield  {journal}
  {\bibinfo  {journal} {Annals of Physics}\ }\textbf {\bibinfo {volume}
  {203}},\ \bibinfo {pages} {231 } (\bibinfo {year} {1990})}\BibitemShut
  {NoStop}%
\bibitem [{\citenamefont {Redlich}(1984{\natexlab{a}})}]{Redlich1984}%
  \BibitemOpen
  \bibfield  {author} {\bibinfo {author} {\bibfnamefont {A.~N.}\ \bibnamefont
  {Redlich}},\ }\bibfield  {title} {\enquote {\bibinfo {title} {{Gauge
  Non-invariance and Parity Non-conservation of Three-Dimensional Fermions}},}\
  }\href {\doibase 10.1103/PhysRevLett.52.18} {\bibfield  {journal} {\bibinfo
  {journal} {Phys. Rev. Lett.}\ }\textbf {\bibinfo {volume} {52}},\ \bibinfo
  {pages} {18} (\bibinfo {year} {1984}{\natexlab{a}})}\BibitemShut {NoStop}%
\bibitem [{\citenamefont {Redlich}(1984{\natexlab{b}})}]{Redlich1984b}%
  \BibitemOpen
  \bibfield  {author} {\bibinfo {author} {\bibfnamefont {A.~N.}\ \bibnamefont
  {Redlich}},\ }\bibfield  {title} {\enquote {\bibinfo {title} {{Parity
  violation and gauge non-invariance of the effective gauge field action in
  three dimensions}},}\ }\href {\doibase 10.1103/PhysRevD.29.2366} {\bibfield
  {journal} {\bibinfo  {journal} {Phys. Rev. D}\ }\textbf {\bibinfo {volume}
  {29}},\ \bibinfo {pages} {2366} (\bibinfo {year}
  {1984}{\natexlab{b}})}\BibitemShut {NoStop}%
\bibitem [{\citenamefont {{Kaul}}\ and\ \citenamefont
  {{Sachdev}}(2008)}]{Kaul2008}%
  \BibitemOpen
  \bibfield  {author} {\bibinfo {author} {\bibfnamefont {R.~K.}\ \bibnamefont
  {{Kaul}}}\ and\ \bibinfo {author} {\bibfnamefont {S.}~\bibnamefont
  {{Sachdev}}},\ }\bibfield  {title} {\enquote {\bibinfo {title} {{Quantum
  criticality of U(1) gauge theories with fermionic and bosonic matter in two
  spatial dimensions}},}\ }\href {\doibase 10.1103/PhysRevB.77.155105}
  {\bibfield  {journal} {\bibinfo  {journal} {Phys. Rev. B}\ }\textbf {\bibinfo
  {volume} {77}},\ \bibinfo {eid} {155105} (\bibinfo {year} {2008})},\ \Eprint
  {http://arxiv.org/abs/0801.0723} {arXiv:0801.0723 [cond-mat.str-el]}
  \BibitemShut {NoStop}%
\bibitem [{\citenamefont {Zucker}(1974)}]{Zucker1974}%
  \BibitemOpen
  \bibfield  {author} {\bibinfo {author} {\bibfnamefont {I.~J.}\ \bibnamefont
  {Zucker}},\ }\bibfield  {title} {\enquote {\bibinfo {title} {Exact results
  for some lattice sums in 2, 4, 6 and 8 dimensions},}\ }\href
  {http://stacks.iop.org/0301-0015/7/i=13/a=011} {\bibfield  {journal}
  {\bibinfo  {journal} {Journal of Physics A: Mathematical, Nuclear and
  General}\ }\textbf {\bibinfo {volume} {7}},\ \bibinfo {pages} {1568}
  (\bibinfo {year} {1974})}\BibitemShut {NoStop}%
\end{thebibliography}%
\end{document}